\documentclass[aps,prd,superscriptaddress]{revtex4}

\usepackage{amsmath}  
\usepackage{amsfonts}
\usepackage{amssymb}
\usepackage{natbib} 
\usepackage{setspace}
\usepackage{graphicx} 
\usepackage{xspace}
\usepackage{cancel}
\usepackage{url}
\usepackage{color}

\def\K0bar{\overline{K^0}}
\def\bge{\begin{equation}}
\def\ene{\end{equation}}
\def\bg{\begin{eqnarray}}
\def\en{\end{eqnarray}}
\def\nn{\nonumber}

\def\qbar{{\bar{q}}}

\def\d0bar{{\bar{D}^0}}

\def\qbar{\bar{q}}
\def\Qbar{\overline{Q}}

\newcommand{\be}{\begin{equation}}
\newcommand{\ee}{\end{equation}}

\begin{document}

\title{
  \vspace{-30mm}
  \begin{flushright} LFTC-18-13/34 \end{flushright}
  \vspace{23mm}
In-medium properties of the low-lying strange, charm, and bottom baryons 
in the quark-meson coupling model
}


\author{K.~Tsushima}
\affiliation{Laborat\'orio de F\'{\i}sica Te\'orica e Computacional-LFTC, 
Universidade Cruzeiro do Sul, 01506-000, S\~ao Paulo, SP, Brazil}


\begin{abstract}
In-medium properties of the low-lying strange, charm, and bottom baryons 
in symmetric nuclear matter are studied in the quark-meson coupling (QMC) model.
Results for the Lorentz-scalar effective masses, mean field potentials felt by the light quarks   
in the baryons, in-medium bag radii, and the lowest mode bag eigenvalues 
are presented for those calculated using the updated data.
This study completes the in-medium properties of the low-lying baryons 
in symmetric nuclear matter in the QMC model, 
for the strange, charm and bottom baryons which contain one or two strange, 
one charm or one bottom quarks, as well as at least one light quark.
Highlight is the prediction of the bottom baryon Lorentz-scalar effective masses,  
namely, the Lorentz-scalar effective mass of $\Sigma_b$ becomes smaller than 
that of $\Xi_b$ at moderate nuclear matter density, $m^*_{\Sigma_b} < m^*_{\Xi_b}$, 
although in vacuum $m_{\Sigma_b} > m_{\Xi_b}$. 
We study further the effects of the repulsive Lorentz-vector potentials on  
the excitation (total) energies of these bottom baryons. 
\end{abstract}



\maketitle

\section{Introduction}

The study of baryon properties in a nuclear medium, especially for   
the baryons which contain charm and/or bottom quarks    
is very interesting~\cite{Jenkins:1990jv,Wise:1992hn,Korner:1994nh,Albertus:2003sx,
Liu:2007fg,Jenkins:2007dm,Brown:2014ena,Simonis:2018rld},    
due to the emergence of heavy-quark symmetry 
also in the baryon sector~\cite{Isgur:1991wq,Mannel:1990vg,Neubert:1993mb}. 
The existence of heavy quarks in hadrons makes it simpler to treat them in many cases,    
e.g., one can treat them in a nonrelativistic framework with effective 
potentials such as nonrelativistic QCD~\cite{Lepage:1992tx,Brambilla:1999xf}. 
In particular, in-medium properties of heavy baryons which contain  
at least one light $u$ or $d$ quarks, can provide us with important 
information on the dynamical chiral symmetry breaking, 
and the roles of light quarks in partial restoration 
of chiral symmetry~\cite{Saito:2005rv,Krein:2017usp,Hosaka:2016ypm}.
Despite of the importance, theoretical studies for the in-medium properties 
of heavy baryons do not seem to exist many~\cite{Hosaka:2016ypm,Azizi:2016dmr,Azizi:2018dtb}, 
probably because the lack of models and/or methods 
which are simple enough to handle easily. 

To study the in-medium properties of heavy baryons,   
we rely here on the quark-meson coupling (QMC) model, 
a quark-based model of nuclear matter, finite nuclei and hadron properties 
in a nuclear medium. The model was invented by Guichon~\cite{Guichon:1987jp}. 
(For other variants of the QMC model, see Ref.~\cite{Saito:2005rv}.)
The QMC model has successfully been applied for various studies   
of the properties of finite (hyper)nuclei~\cite{Guichon:1995ue,Saito:1996sf,
Saito:1996yb,Stone:2016qmi,Guichon:2018uew,Tsushima:1997cu,
Tsushima:1997rd,Guichon:2008zz,Tsushima:2002ua,Tsushima:2002sm,
Tsushima:2003dd}, hadron properties 
in a nuclear medium~\cite{Saito:1997ae,Tsushima:1997df,
Tsushima:1998qw,Tsushima:1998ru,Sibirtsev:1999jr,Tsushima:2002cc}, 
reactions involving nuclear targets~\cite{Sibirtsev:1999js,Shyam:2008ny,
Tsushima:2009zh,Shyam:2011aa,Chatterjee:2012ja,Tsushima:2012pt,Shyam:2016bzq,
Shyam:2016uxa,Shyam:2018iws}, and neutron star 
structure~\cite{Whittenbury:2013wma,Thomas:2013sea}.
Self-consistent exchange of the Lorentz-scalar-isoscalar $\sigma$-,  
Lorentz-vector-isoscalar $\omega$-, and Lorentz-vector-isovector $\rho$-mean fields,  
directly couple to the light quarks $u$ and $d$, is the key feature of the model   
to be able to achieve the novel saturation properties of nuclear matter  
with a simple and systematic treatment. 
All the relevant coupling constants of the $\sigma$-light-quark,  
$\omega$-light-quark, and $\rho$-light-quark in any hadrons, 
are the same as those in nucleon, 
those fixed by the nuclear matter saturation properties. 
The physics behind of this simple picture may be supported by the fact 
that the light-quark condensates reduces/changes faster than those of 
the strange and heavier quarks in finite 
density as the nuclear density increases~\cite{Tsushima:1991fe,Maruyama:1992ab}.  
Or, partial restoration of chiral symmetry in nuclear medium 
is mainly driven by the decrease in the magnitude of the light quark condensates.
This is modeled in the QMC model by the fact that  
the scalar-isoscalar $\sigma$-, vector-isoscalar $\omega$-, and vector-isovector $\rho$-mean 
fields couple directly only to the light quarks, but not to the strange 
nor heavier quarks.

The present article completes the studies for  
the low-lying baryon properties in symmetric nuclear matter 
in the QMC model with some updates.
In particular, highlight is on the bottom baryon Lorentz-scalar effective masses in nuclear medium.    
Detailed results are presented explicitly, where many of them have not been presented 
before~\cite{Saito:2005rv,Krein:2017usp}.

We predict that the Lorentz-scalar effective mass of $\Sigma_b$
becomes smaller than that of $\Xi_b$ at moderate nuclear matter density,     
namely, $m^*_{\Sigma_b} < m^*_{\Xi_b}$, 
although $m_{\Sigma_b} > m_{\Xi_b}$ in vacuum. 
We study further the effects of the repulsive Lorentz-vector potentials on the 
excitation (total) energies of these bottom baryons, by considering two different  
possibilities for the vector potentials,  
one is extracted by the $\Lambda$ and $\Sigma$ hypernucear 
experimental observation, the one which includes effective 
{\it Pauli potentials} based on the Pauli-principle 
at the quark level, and the other is the vector potentials  
that are predicted by the QMC model without the effective {\it Pauli potentials}.

\section{Finite (hyper)nucleus in the QMC model}
\label{qmc}

In order to make this article self-contained, we briefly review  
the QMC model following Ref.~\cite{Saito:2005rv,Krein:2017usp}  
with minor improvements for better understanding.

Although Hartree-Fock treatment is possible within the QMC model~\cite{Krein:1998vc}, 
the main features of the results, especially the density dependence of 
total energy per nucleon (nuclear matter energy density) is nearly identical as that of 
the Hartree approximation. Then,  
it is sufficient to rely on the Hartree approximation in this study. 
(See Ref.~\cite{Whittenbury:2013wma} for a detailed study made for the neutron star structure 
based on the QMC model with the Hartree-Fock treatment.)

Before discussing the heavy baryon properties in symmetric nuclear matter, 
we start by the case of finite (hyper)nucleus. 
Using the Born-Oppenheimer approximation, 
a relativistic Lagrangian density which gives the same mean-field equations
of motion for a nucleus or a hypernucleus, may be given in 
the QMC model~\cite{Saito:2005rv,Krein:2017usp,Tsushima:1997cu} below, 
where the quasi-particles moving in single-particle orbits are three-quark 
clusters with the quantum numbers of a nucleon, strange, charm or bottom hyperon 
when expanded to the same order in 
velocity~\cite{Guichon:1995ue,Saito:1996sf,Tsushima:1997cu,Tsushima:2002ua,
Tsushima:2003dd,Tsushima:2002cc}:  
\begin{eqnarray}
{\cal L}^{Y}_{QMC} &=& {\cal L}^N_{QMC} + {\cal L}^Y_{QMC},
\label{eq:LagYQMC} \\
{\cal L}^N_{QMC} &\equiv&  \overline{\psi}_N(\vec{r})
\left[ i \gamma \cdot \partial
- m_N^*(\sigma) - (\, g_\omega \omega(\vec{r})
+ g_\rho \dfrac{\tau^N_3}{2} b(\vec{r})
+ \dfrac{e}{2} (1+\tau^N_3) A(\vec{r}) \,) \gamma_0
\right] \psi_N(\vec{r}) \quad \nn \\
  & & - \dfrac{1}{2}[ (\nabla \sigma(\vec{r}))^2 +
m_{\sigma}^2 \sigma(\vec{r})^2 ]
+ \dfrac{1}{2}[ (\nabla \omega(\vec{r}))^2 + m_{\omega}^2
\omega(\vec{r})^2 ] \nn \\
 & & + \dfrac{1}{2}[ (\nabla b(\vec{r}))^2 + m_{\rho}^2 b(\vec{r})^2 ]
+ \dfrac{1}{2} (\nabla A(\vec{r}))^2, \label{eq:LagN} \\
{\cal L}^Y_{QMC} &\equiv&
\overline{\psi}_Y(\vec{r})
\left[ i \gamma \cdot \partial
- m_Y^*(\sigma)
- (\, g^Y_\omega \omega(\vec{r})
+ g^Y_\rho I^Y_3 b(\vec{r})
+ e Q_Y A(\vec{r}) \,) \gamma_0
\right] \psi_Y(\vec{r}), 
\nn\\
& &\hspace{30ex} (Y = \Lambda,\Sigma^{0,\pm},\Xi^{0,-},
\Lambda^+_c,\Sigma_c^{0,+,++},\Xi_c^{0,+},\Lambda_b,\Sigma_b^{0,\pm},\Xi_b^{0,-}),
\label{eq:LagY}
\end{eqnarray}
where, for a normal nucleus, ${\cal L}^Y_{QMC}$ in Eq.~(\ref{eq:LagYQMC}), 
namely Eq.~(\ref{eq:LagY}) is not needed, but for the following study we do need this.
In the above $\psi_N(\vec{r})$ and $\psi_Y(\vec{r})$
are respectively the nucleon and hyperon (strange, charm or bottom baryon) fields. 
The mean-meson fields represented by, $\sigma, \omega$ and $b$ which 
directly couple to the light quarks self-consistently, are  
the Lorentz-scalar-isoscalar, Lorentz-vector-isoscalar and third component of  
Lorentz-vector-isovector fields, respectively, while $A$ stands for the Coulomb field.

In an approximation where the $\sigma$-, $\omega$- and $\rho$-mean fields couple
only to the $u$ and $d$ light quarks, the coupling constants for the hyperon 
appearing in Eq.~(\ref{eq:LagY})  
are obtained/identified as $g^Y_\omega = (n_q/3) g_\omega$, and
$g^Y_\rho \equiv g_\rho = g_\rho^q$, with $n_q$ being the total number of
valence light quarks in the hyperon $Y$, where $g_\omega$ and $g_\rho$ are 
the $\omega$-$N$ and $\rho$-$N$ coupling constants. $I^Y_3$ and $Q_Y$
are the third component of the hyperon isospin operator and its electric
charge in units of the proton charge, $e$, respectively.

As mentioned already, the approximation adopted in the QMC model, that the meson fields couple 
only to the light quarks, reflects the fact 
that the magnitudes of the light-quark condensates decrease faster 
as increasing the nuclear density than those of the strange and heavy flavor quarks. 
This is associated with partial restoration of chiral symmetry 
in a nuclear medium (dynamically symmetry breaking 
and its partial restoration). 
The dynamical symmetry breaking and its restoration can provide us 
with important information on the origin of the (dynamical) masses 
of hadrons which we observe in our universe.

The field dependent $\sigma$-$N$ and $\sigma$-$Y$
coupling strengths respectively for the nucleon $N$ and hyperon $Y$,  
$g_\sigma(\sigma) \equiv g^N_\sigma(\sigma)$ and  $g^Y_\sigma(\sigma)$ 
appearing in Eqs.~(\ref{eq:LagN}) and~(\ref{eq:LagY}), are defined by
\bg
& &m_N^*(\sigma) \equiv m_N - g_\sigma(\sigma)
\sigma(\vec{r}),  
\label{effnmass}
\\
& &m_Y^*(\sigma) \equiv m_Y - g^Y_\sigma(\sigma)
\sigma(\vec{r}) 
\hspace{3ex} (Y = \Lambda,\Sigma,\Xi, 
\Lambda_c,\Sigma_c,\Xi_c,\Lambda_b,\Sigma_b,\Xi_b), 
\label{effymass}
\en
where $m_N$ ($m_Y$) is the free nucleon (hyperon) mass. 
Note that the dependence of these coupling strengths on the applied
scalar field ($\sigma$) must be calculated self-consistently within the quark
model~\cite{Guichon:1987jp,Guichon:1995ue,Tsushima:1997cu,Tsushima:2002ua,
Tsushima:2002sm,Tsushima:2002cc}.
Hence, unlike quantum hadrodynamics (QHD)~\cite{Walecka:1974qa,Serot:1984ey}, even though
$g^Y_\sigma(\sigma) / g_\sigma(\sigma)$ may be
2/3 or 1/3 depending on the number of light quarks $n_q$ in the hyperon 
in free space, $\sigma = 0$ (even this is true only when their bag 
radii in free space are exactly the same in the standard QMC model with the MIT bag), 
this will not necessarily be the case in a nuclear medium.

The Lagrangian density Eq.~(\ref{eq:LagYQMC}) [or (\ref{eq:LagN}) and (\ref{eq:LagY})] 
leads [lead] to a set of equations of motion for the finite (hyper)nuclear system:
\begin{eqnarray}
& &[i\gamma \cdot \partial -m^*_N(\sigma)-
(\, g_\omega \omega(\vec{r}) + g_\rho \dfrac{\tau^N_3}{2} b(\vec{r})
 + \dfrac{e}{2} (1+\tau^N_3) A(\vec{r}) \,)
\gamma_0 ] \psi_N(\vec{r}) = 0, \label{eqdiracn}\\
& &[i\gamma \cdot \partial - m^*_Y(\sigma)-
(\, g^Y_\omega \omega(\vec{r}) + g_\rho I^Y_3 b(\vec{r})
+ e Q_Y A(\vec{r}) \,)
\gamma_0 ] \psi_Y(\vec{r}) = 0, \label{eqdiracy}\\
& &(-\nabla^2_r+m^2_\sigma)\sigma(\vec{r}) =
- \left[\dfrac{\partial m_N^*(\sigma)}{\partial \sigma}\right]\rho_s(\vec{r})
- \left[\dfrac{\partial m_Y^*(\sigma)}{\partial \sigma}\right]\rho^Y_s(\vec{r}),
\nn \\
& & \hspace{7.5em} \equiv g_\sigma C_N(\sigma) \rho_s(\vec{r})
    + g^Y_\sigma C_Y(\sigma) \rho^Y_s(\vec{r}) , \label{eqsigma}\\
& &(-\nabla^2_r+m^2_\omega) \omega(\vec{r}) =
g_\omega \rho_B(\vec{r}) + g^Y_\omega
\rho^Y_B(\vec{r}) ,\label{eqomega}\\
& &(-\nabla^2_r+m^2_\rho) b(\vec{r}) =
\dfrac{g_\rho}{2}\rho_3(\vec{r}) + g^Y_\rho I^Y_3 \rho^Y_B(\vec{r}),
 \label{eqrho}\\
& &(-\nabla^2_r) A(\vec{r}) =
e \rho_p(\vec{r})
+ e Q_Y \rho^Y_B(\vec{r}) ,\label{eqcoulomb}
\end{eqnarray}
where, $\rho_s(\vec{r})$ ($\rho^Y_s(\vec{r})$), $\rho_B(\vec{r})=\rho_p(\vec{r})+\rho_n(\vec{r})$
($\rho^Y_B(\vec{r})$), $\rho_3(\vec{r})=\rho_p(\vec{r})-\rho_n(\vec{r})$, 
$\rho_p(\vec{r})$ and $\rho_n(\vec{r})$ are the nucleon (hyperon) scalar, 
nucleon (hyperon) baryon, third component of isovector,
proton and neutron densities at the position $\vec{r}$ in
the (hyper)nucleus. On the right hand side of Eq.~(\ref{eqsigma}),
$- [{\partial m_N^*(\sigma)}/{\partial \sigma}] \equiv
g_\sigma C_N(\sigma)$ and
$- [{\partial m_Y^*(\sigma)}/{\partial \sigma}] \equiv
g^Y_\sigma C_Y(\sigma)$, where $g_\sigma \equiv g_\sigma (\sigma=0)$ and
$g^Y_\sigma \equiv g^Y_\sigma (\sigma=0)$ hereafter all in this article, 
are the key ingredients of the QMC model. 
Note that, when there is $\sigma$-dependence, they will be explicitly written by  
$g_\sigma(\sigma)$ and $g^Y_\sigma(\sigma)$ to avoid confusion. 
At the hadronic level, the entire information
of the quark dynamics is condensed in the effective couplings
$C_{N,Y}(\sigma)$ of Eq.~(\ref{eqsigma}), which characterize the  
features of the QMC model, namely, {\it scalar polarisability}. 
Furthermore, when $C_{N,Y}(\sigma) = 1$, which corresponds to
a structureless nucleon or hyperon, the equations of motion
given by Eqs.~(\ref{eqdiracn})-(\ref{eqcoulomb})
can be identified with those derived from naive QHD~\cite{Walecka:1974qa,Serot:1984ey}.

We note that, for the Dirac equation Eq.~(\ref{eqdiracy})
for the hyperon $Y$,  
we include the effects due to the Pauli blocking 
at the quark level by adding repulsive potentials based on 
the study made for the strange hyperons $\Lambda, \Sigma$, and $\Xi$. 
The net, repulsive ``Pauli potentials'', which may be interpreted as also including 
the $\Lambda N - \Sigma N$ channel coupling effect,  
was extracted by the fit to the $\Lambda$- and $\Sigma$-hypernuclei taking 
into account the $\Sigma N - \Lambda N$ channel coupling~\cite{Tsushima:1997cu}. 
Of course, the effects of the channel coupling are expected to be smaller for the 
corresponding charm and bottom baryons, since the corresponding mass differences for 
these cases are larger than 
that for the $\Lambda$ and $\Sigma$ hyperons. 
Thus, for the interesting case of the $\Sigma_b - \Xi_b$ baryon system focused on later,   
we study two possibilities of the vector potentials, 
with and without including the effective Pauli potentials.
The modified Dirac equation for  $Y = \Lambda, \Sigma, \Xi, \Lambda_{c,b}, \Sigma_{c,b}$ 
and $\Xi_{c,b}$ is, 
\begin{equation}
 [i\gamma \cdot \partial - M_Y(\sigma)-
(\, \lambda_Y \rho_B(\vec{r}) + g^Y_\omega \omega(\vec{r}) 
+ g_\rho I^Y_3 b(\vec{r}) 
 + e Q_Y A(\vec{r}) \,) \gamma_0 ] \psi_Y(\vec{r}) = 0,
\label{Pauli}
\end{equation}
where $\lambda_Y \rho_B(\vec{r})$ is the effective Pauli potential 
for the hyperon $Y$, with $\rho_B(\vec{r})$ being
the baryon density at the position $\vec{r}$ in the corresponding hypernucleus. 
The values of $\lambda_Y$ for $Y = (\Lambda,\Lambda_{c,b}$), and $(\Sigma,\Sigma_{c,b})$ are  
respectively 60.25 MeV (fm)$^3$ and 110.6 MeV (fm)$^3$, while for $Y = \Xi$ and $\Xi_{c,b}$,   
$\lambda_Y$ is $(1/2) \times 60.25$ MeV (fm)$^3$ based on the valence light-quark number. 
For the details of the effective Pauli potentials at 
the quark level, see Ref.~\cite{Tsushima:1997cu}.

The effective masses of the nucleon $N$ ($m^*_N$) 
and hyperon $Y$ ($m^*_Y$) are calculated later by Eq.~(\ref{hmass}) 
(by replacing $h \to N$, and $h \to Y$, respectively there).
The explicit expressions for 
$C_{N,Y}(\sigma) \equiv S_{N,Y}(\sigma) / S_{N,Y}(\sigma=0)$ 
($S_{N,Y}(\sigma)$ to be defined next) 
and the effective masses $m^*_{N,Y}$ are related by,
\bg
\dfrac{\partial m_{N,Y}^*(\sigma)}{\partial \sigma}
&=& - n_q g_{\sigma}^q \int_{bag} d^3y 
{\overline \psi}_q(\vec{y}) \psi_q(\vec{y})
\nn\\
&\equiv& - n_q g_\sigma^q S_{N,Y}(\sigma) 
= - \left[ n_q g_\sigma^q S_{N,Y} (\sigma=0)\right]\, C_{N,Y}(\sigma) 
= - \dfrac{\partial}{\partial \sigma}
\left[ g^{N,Y}_\sigma(\sigma) \sigma \right],
\label{Ssigma}
\en
where $g^q_\sigma$ is the light-quark-$\sigma$ coupling constant, 
and $\psi_q$ is the light quark wave function in the nucleon $N$ or 
hyperon $Y$ immersed in a nuclear medium.
By the above relation, we define the $\sigma$-$N$ and $\sigma$-$Y$ coupling 
constants,
\begin{equation}
g^{N,Y}_\sigma \equiv n_q g^q_\sigma S_{N,Y} (\sigma = 0), 
\label{sigmacc}
\end{equation}
where $g^N_\sigma \equiv g_\sigma = g_\sigma(\sigma=0)$ appeared already. 
Note that, as in the case of $C_{N,Y}(\sigma)$, the values of 
$S_N(\sigma = 0)$ and $S_Y(\sigma = 0)$  
are different, because the light-quark wave functions in the nucleon $N$ and hyperon $Y$ 
are different in vacuum as well as in medium; that is, the bag radii of the $N$ and $Y$ 
are different in both vacuum and medium.

The parameters appearing at the nucleon, hyperon and meson Lagrangian level 
used for the study of infinite nuclear matter and finite 
nuclei~\cite{Guichon:1995ue,Saito:1996sf} are: 
$m_\omega = 783$ MeV, $m_\rho = 770$ MeV, $m_\sigma = 550$ MeV and $e^2/4\pi = 1/137.036$.
(See Ref.~\cite{Saito:1996sf}
for a discussion on the parameter fixing in the QMC model, 
especially in treating finite nuclei.)

\section{Baryon properties in symmetric nuclear matter}
\label{matter}

We consider the rest frame of infinitely large,  
symmetric nuclear matter, a spin and isospin saturated system 
with only strong interaction (Coulomb force is dropped as usual).
One first keeps only ${\cal L}^N_{QMC}$ in Eq.~(\ref{eq:LagYQMC}), 
or correspondingly drops all the quantities with the super- and under-scripts $Y$, 
and sets the Coulomb field $A(\vec{r})=0$ in Eqs.~(\ref{eqdiracn})-(\ref{eqcoulomb}). 
Next one sets all the terms with any derivatives of the fields to be zero.  
Then, within the Hartree mean-field approximation, 
the nuclear (baryon) $\rho_B$, and scalar $\rho_s$ densities 
are respectively given by,
\begin{eqnarray}
\rho_B &=& \dfrac{4}{(2\pi)^3}\int d^3{k}\ \theta (k_F - |\vec{k}|)
= \dfrac{2 k_F^3}{3\pi^2},
\label{rhoB}
\\
\rho_s &=& \dfrac{4}{(2\pi)^3}\int d^3 {k} \ \theta (k_F - |\vec{k}|)
\dfrac{m_N^*(\sigma)}{\sqrt{m_N^{* 2}(\sigma)+\vec{k}^2}} \, .
\label{rhos}
\end{eqnarray}
Here, $m^*_N(\sigma)$ is the value (constant) of the Lorentz-scalar effective nucleon mass at 
a given nuclear (baryon) density 
(see also Eq.~(\ref{effnmass})) and $k_F$ the Fermi momentum.
In the standard QMC model~\cite{Guichon:1987jp}, the MIT bag model is used 
for describing nucleons and hyperons (hadrons). The use of this quark model is 
an essential ingredient for the QMC model, namely the use of the relativistic, 
confined quarks.

The Dirac equations for the quarks and antiquarks 
in nuclear matter, in a bag of a hadron, $h$, ($q = u$ or $d$, and $Q = s,c$ or $b$, hereafter) 
neglecting the Coulomb force, are given by  
( $x=(t,\vec{x})$ and for $|\vec{x}|\le$ 
bag radius)~\cite{Tsushima:1997df,Tsushima:1998ru,Sibirtsev:1999jr,Tsushima:2002cc,Sibirtsev:1999js}, 
\begin{eqnarray}
\left[ i \gamma \cdot \partial_x -
(m_q - V^q_\sigma)
\mp \gamma^0
\left( V^q_\omega +
\dfrac{1}{2} V^q_\rho
\right) \right] 
\left( \begin{array}{c} \psi_u(x)  \\
\psi_{\bar{u}}(x) \\ \end{array} \right) &=& 0,
\label{Diracu}\\
\left[ i \gamma \cdot \partial_x -
(m_q - V^q_\sigma)
\mp \gamma^0
\left( V^q_\omega -
\dfrac{1}{2} V^q_\rho
\right) \right]
\left( \begin{array}{c} \psi_d(x)  \\
\psi_{\bar{d}}(x) \\ \end{array} \right) &=& 0,
\label{Diracd}\\
\left[ i \gamma \cdot \partial_x - m_{Q} \right] \psi_{Q} (x) = 0, ~~~~~~~~  
\left[ i \gamma \cdot \partial_x - m_{Q} \right] \psi_{\overline{Q}} (x) &=& 0,  
\label{DiracQ}
\end{eqnarray}
where, the (constant) mean fields for a bag in nuclear matter
are defined by $V^q_\sigma \equiv g^q_\sigma \sigma$, 
$V^q_\omega \equiv g^q_\omega \omega$ and
$V^q_\rho \equiv g^q_\rho b$,
with $g^q_\sigma$, $g^q_\omega$ and
$g^q_\rho$ being the corresponding quark-meson coupling constants. 
We assume SU(2) symmetry, $m_{u,\bar{u}}=m_{d,\bar{d}} \equiv m_{q,\bar{q}}$. 
The corresponding Lorentz-scalar effective quark masses are defined  
by, $m^*_{u,\bar{u}}=m^*_{d,\bar{d}}=m^*_{q,\bar{q}} \equiv m_{q,\bar{q}}-V^q_{\sigma}$.
Since the $\rho$-meson mean field becomes zero, $V^q_{\rho}=0$~in Eqs.~(\ref{Diracu}) 
and~(\ref{Diracd}) in symmetric nuclear matter in the Hartree approximation,      
we will ignore it. (This is not true in a finite nucleus 
with equal and more than two protons  
even with equal numbers of protons and neutrons, 
since the Coulomb interactions among the protons induce an asymmetry  
between the proton and neutron density distributions 
to give $\rho_3(\vec{r}) = \rho_p(\vec{r}) - \rho_n(\vec{r}) \ne 0$.)

The same meson-mean fields $\sigma$ and $\omega$ for the quarks  
in Eqs.~(\ref{Diracu}) and~(\ref{Diracd}), satisfy self-consistently 
the following equations at the nucleon level (together with the Lorentz-scalar effective 
nucleon mass $m_N^*(\sigma)$ of Eq.~(\ref{effnmass}) 
to be calculated by Eq.~(\ref{hmass})):
\begin{eqnarray}
{\omega}&=&\dfrac{g_\omega}{m_\omega^2} \rho_B,
\label{omgf}\\
{\sigma}&=&\dfrac{g_\sigma }{m_\sigma^2}C_N({\sigma})
\dfrac{4}{(2\pi)^3}\int d^3{k} \ \theta (k_F - |\vec{k}|)
\dfrac{m_N^*(\sigma)}{\sqrt{m_N^{* 2}(\sigma)+\vec{k}^2}}
=\dfrac{g_\sigma }{m_\sigma^2}C_N({\sigma}) \rho_s,
\label{sigf}
\end{eqnarray}
where 
\be
C_N(\sigma) \equiv \dfrac{-1}{g_\sigma(\sigma=0)}
\left[ {\partial m^*_N(\sigma)}/{\partial\sigma} \right].
\ee
Because of the underlying quark structure of the nucleon used to calculate
$m^*_N(\sigma)$ in nuclear medium, $C_N(\sigma)$ decreases as $\sigma$ increases,
whereas in the usual point-like nucleon-based models it is constant, $C_N(\sigma) = 1$. 
As will be discussed later it can be parametrized in the QMC model as 
$C_N(\sigma) = 1- a_N \times (g_\sigma \sigma)\,  (a_N > 0)$.
It is this variation of $C_N(\sigma)$ (or equivalently dependence of the scalar coupling on density,
or $\sigma$, $g_\sigma (\sigma)$) that yields a novel saturation mechanism for nuclear matter 
in the QMC model, and contains the important dynamics which originates from the quark structure
of the nucleons and hadrons. It is the variation of this $C_N(\sigma)$, 
which yields three-body or density dependent effective forces, as has been demonstrated 
by constructing an equivalent energy density functional~\cite{Guichon:2018uew,Guichon:2006er}.   
As a consequence of the {\em derived}, nonlinear
couplings of the meson fields in the Lagrangian density at the nucleon (hyperon) and meson level,
the standard QMC model yields the nuclear incompressibility of $K \simeq 280$~MeV with 
$m_q=5$ MeV. This is in contrast to a naive version of QHD~\cite{Walecka:1974qa,Serot:1984ey}
(the point-like nucleon model of nuclear matter),
results in the much larger value, $K \simeq 500$~MeV;
the empirically extracted value falls in the range $K = 200 - 300$ MeV.
(See Ref.~\cite{Dutra:2012mb} for an extensive analysis on this issue.)

%
\begin{table}[htb]
\begin{center}
\caption{
Current quark mass values (inputs), quark-meson coupling constants 
and the bag pressure, $B_p$. Note that the $m_c$ value is 
updated from Refs.~\cite{Saito:2005rv,Krein:2017usp} 
based on the data~\cite{PDG}. 
}
\label{coupcc}
\vspace{1ex}
\begin{tabular}[t]{r|r||l|l}
\hline
\hline
$m_{u,d}$ &5    MeV &$g^q_\sigma$ &5.69\\
$m_s$     &250  MeV &$g^q_\omega$ &2.72\\
$m_c$     &1270 MeV &$g^q_\rho$   &9.33\\
$m_b$     &4200 MeV &$B_p^{1/4}$    &170 MeV\\
\hline
\hline
\end{tabular}
\end{center}
\end{table}

\begin{figure}[htb]
\vspace{4ex}
\centering
\includegraphics[scale=0.3]{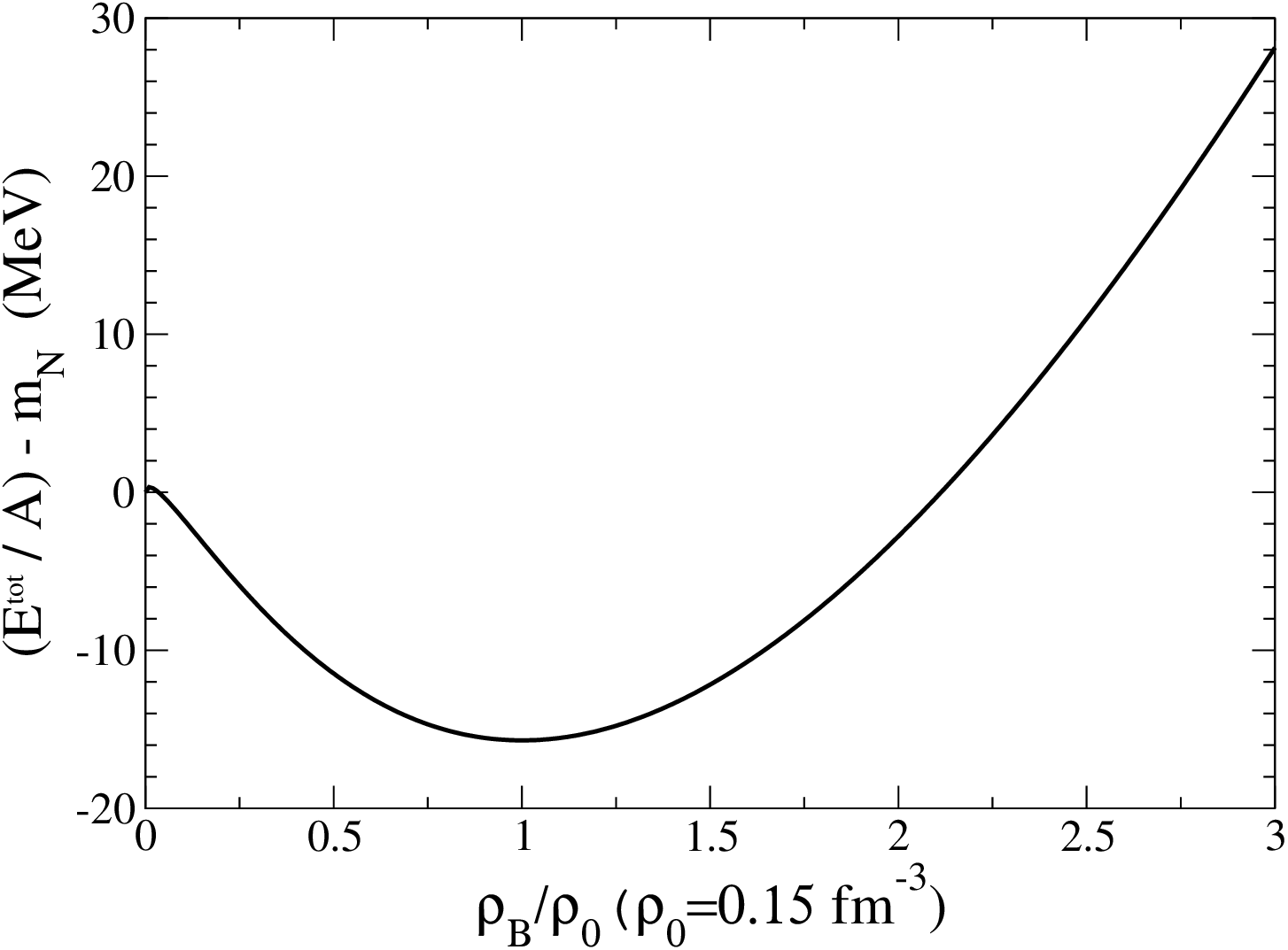}
\hspace{1ex}
\includegraphics[scale=0.3]{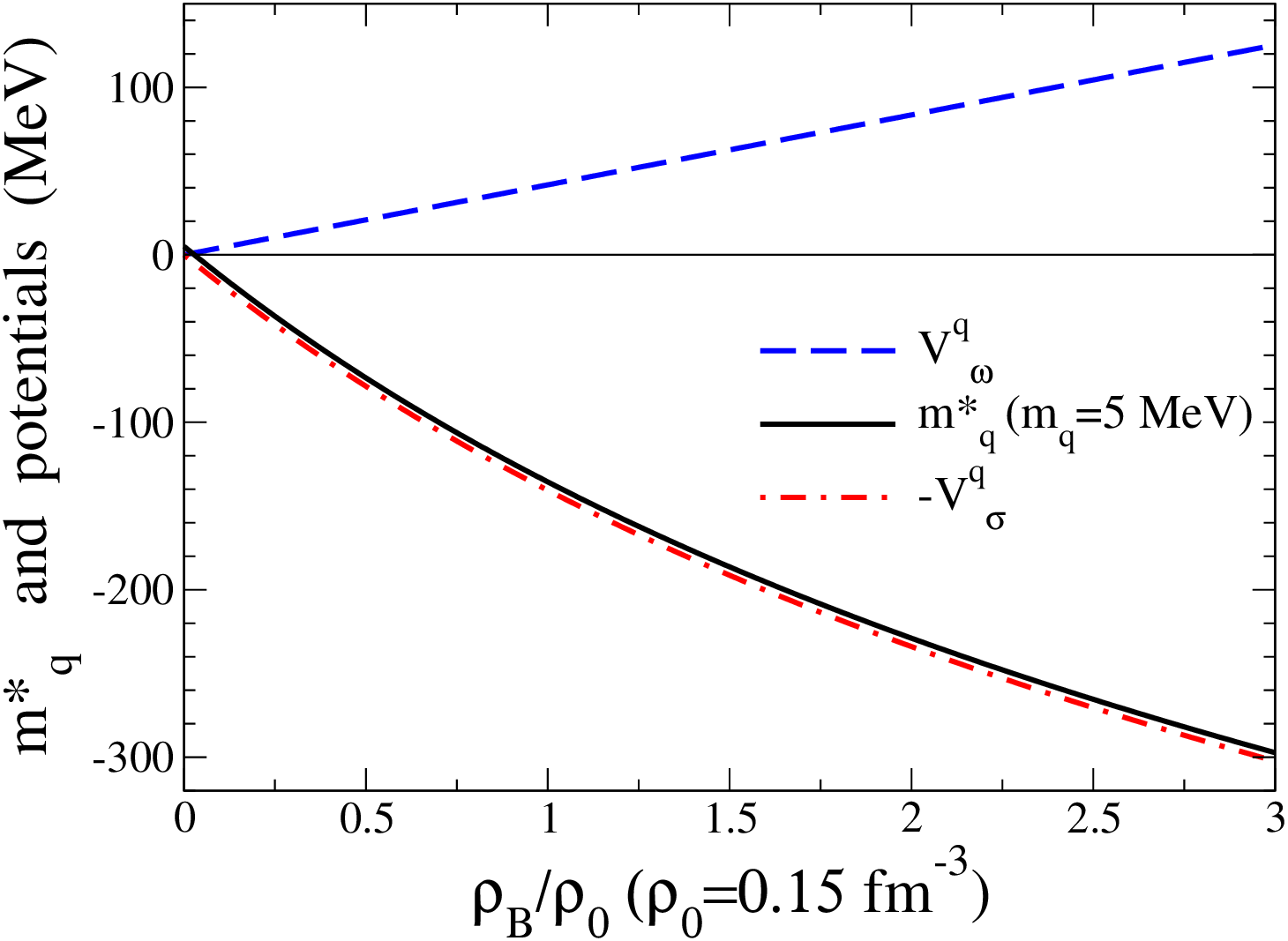}
\caption{\label{fig:Edenpot}
Total energy per nucleon $E^{\rm tot}/A - m_N$ (left panel), and 
the light-quark Lorentz-scalar effective mass $m^*_q$, vector ($V^q_\omega$) and 
scalar ($-V^q_\sigma$) potentials felt by the light quarks.
}
\vspace{2ex}
\end{figure}

Once the self-consistency equation for the ${\sigma}$ field 
Eq.~(\ref{sigf}) is solved, one can evaluate the total energy per nucleon:
\begin{equation}
E^\mathrm{tot}/A=\dfrac{4}{(2\pi)^3 \rho_B}\int d^3{k} \
\theta (k_F - |\vec{k}|) \sqrt{m_N^{* 2}(\sigma)+
\vec{k}^2}+\dfrac{m_\sigma^2 {\sigma}^2}{2 \rho_B}+
\dfrac{g_\omega^2 \rho_B}{2m_\omega^2} .
\label{toten}
\end{equation}
We then determine the coupling constants, $g_{\sigma}$ and $g_{\omega}$
at the nucleon level (see also Eq.~(\ref{sigmacc})), by 
the fit to the binding energy of 15.7~MeV at the saturation density $\rho_0$ = 0.15 fm$^{-3}$
($k_F^0$ = 1.305 fm$^{-1}$) for symmetric nuclear matter, as well as  
$g_\rho$ to the symmetry energy of~35 MeV.
The determined quark-meson coupling constants, and the current quark mass values 
are listed in table~\ref{coupcc}.
The coupling constants at the nucleon level are  
$g^2_\sigma/4\pi = 3.12$, $g^2_\omega/4\pi = 5.31$ and $g^2_\rho/4\pi = 6.93$.
(See Eq.~(\ref{sigmacc}) for $g_\sigma = g^N_\sigma$.)

We show in Fig.~\ref{fig:Edenpot} the density dependence of the total 
energy per nucleon  
$E^{\rm tot}/A - m_N$ (left panel) and the Lorentz-scalar effective quark mass $m^*_q$, 
vector ($V^q_\omega$) and scalar ($-V^q_\sigma$) potentials 
felt by the light quarks (right panel)  
calculated using the quark-meson coupling constants determined.

In the following, let us consider the situation that a hadron $h$ (or a hyperon $Y$) is immersed 
in nuclear matter. The normalized, static solution for the ground state quarks or antiquarks
with flavor $f$ in the hadron $h$ may be written,  
$\psi_f (x) = N_f \exp^{- i \epsilon_f t / R_h^*} \psi_f (\vec{r})$,
where $N_f$ and $\psi_f(\vec{r})$ are the normalization factor and
corresponding spin and spatial part of the wave function. 
The bag radius in medium for the hadron $h$, denoted by $R_h^*$, 
is determined through the
stability condition for the mass of the hadron against the
variation of the bag 
radius~\cite{Guichon:1987jp,Tsushima:1997rd}
(see Eq.~(\ref{hmass})). 
The eigenenergies in units of $1/R_h^*$ are given by, 
\bge
\left( \begin{array}{c}
\epsilon_u \\
\epsilon_{\bar{u}}
\end{array} \right)
= \Omega_q^* \pm R_h^* \left(
V^q_\omega
+ \dfrac{1}{2} V^q_\rho \right),\,\,
\left( \begin{array}{c} \epsilon_d \\
\epsilon_{\bar{d}}
\end{array} \right)
= \Omega_q^* \pm R_h^* \left(
V^q_\omega
- \dfrac{1}{2} V^q_\rho \right),\,\,
\epsilon_{Q}
= \epsilon_{\Qbar} =
\Omega_{Q}.
\label{energy}
\ene

\noindent
The hadron mass in a nuclear medium, $m^*_h$ (free mass is denoted by $m_h$), is calculated by
\begin{eqnarray}
m_h^* &=& \sum_{j=q,\bar{q},Q,\Qbar} 
\dfrac{ n_j\Omega_j^* - z_h}{R_h^*}
+ \frac{4}{3}\pi R_h^{* 3} B_p,\quad
\left. \dfrac{\partial m_h^*}
{\partial R_h}\right|_{R_h = R_h^*} = 0,
\label{hmass}
\end{eqnarray}
where $\Omega_q^*=\Omega_{\bar{q}}^*
=[x_q^2 + (R_h^* m_q^*)^2]^{1/2}\,(q=u,d)$, with
$m_q^*=m_q{-}g^q_\sigma \sigma=m_q-V^q_\sigma$,
$\Omega_Q^*=\Omega_{\Qbar}^*=[x_Q^2 + (R_h^* m_Q)^2]^{1/2}\,(Q=s,c,b)$,
and $x_{q,Q}$ are the lowest mode bag eigenvalues.
$B_p$ is the bag pressure (constant), $n_q (n_{\qbar})$ and $n_Q (n_{\Qbar})$ 
are the lowest mode valence quark (antiquark) 
numbers for the quark flavors $q$ and $Q$ 
in the hadron $h$, respectively, 
while $z_h$ parametrizes the sum of the
center-of-mass and gluon fluctuation effects,   
which are assumed to be
independent of density~\cite{Guichon:1995ue}. 
The bag pressure $B_p = {\rm (170\, MeV)}^4$ (density independent) is determined 
by the free nucleon mass 
$m_N = 939$ MeV with the bag radius in vacuum $R_N = 0.8$ fm and $m_q = 5$ MeV as inputs, 
which are considered to be standard values in the QMC model~\cite{Saito:2005rv}.
(See also table~\ref{coupcc}.)
Concerning the Lorentz-scalar effective mass $m_q^*$ in nuclear 
medium, it reflects nothing but the strength 
of the attractive scalar potential 
as in Eqs.~(\ref{Diracu}) and~(\ref{Diracd}), 
and thus naive interpretation of the mass for a (physical) particle, 
which is positive, should not be applied. 
The model parameters are determined to reproduce the corresponding masses in free space.
The quark-meson coupling constants, $g^q_\sigma$, $g^q_\omega$
and $g^q_\rho$, have already been determined by the nuclear matter saturation properties.
Exactly the same coupling constants, $g^q_\sigma$, $g^q_\omega$ and
$g^q_\rho$, will be used for the light quarks in all the hadrons  
as in the nucleon. These values are fixed, and will not be changed 
depending on the hadrons.

In table~\ref{bagparambc} we present the inputs, vacuum masses of baryons $B$, $m_B$,   
the parameters $z_B$, the calculated lowest mode bag eigenvalues ($x_1,x_2,x_3$) 
of the corresponding valence quarks ($q_1,q_2,q_3$) in the baryon $B$,  
and the bag radii calculated in vacuum $R_B$, as well as the corresponding quantities at 
$\rho_0 = 0.15$ fm$^{-3}$, namely, the Lorentz-scalar effective masses $m^*_B$, 
in-medium bag radii $R^*_B$,  
and the lowest mode bag eigenvalues, ($x^*_1,x^*_2,x^*_3$). 
Note that in the QMC model, $\Omega(sss), \Omega_c(ssc)$ and 
$\Omega_b(ssb)$ properties are not modified in medium.

\begin{table}
\begin{center}
\caption{
The parameters related with the zero-point energy $z_B$,
baryon masses and the bag radii in free space
[at normal nuclear matter density, $\rho_0=0.15$ fm$^{-3}$]
$m_B$(MeV), $R_B$(fm) [$m^*_B, R^*_B$], the lowest mode bag eigenvalues $x_1, x_2, x_3$ 
[$x^*_1, x^*_2, x^*_3$] of baryon $B(q_1,q_2,q_3)$ with 
the corresponding valence quarks $q_1, q_2, q_3$ in the baryon $B$, 
where $z_B$s are kept the same as those in vacuum, i.e., density independent.
Free space mass values $m_B$ for the heavy baryons are from Ref.~\cite{PDG},   
and those for the strange hyperons are from Ref.~\cite{Saito:2005rv}, 
as well as the nucleon bag radius $R_N = 0.8$ fm (and $m_q=5$ MeV), are inputs.
The light quarks are indicated by $q = u$ or $d$.
Note that, the baryons containing at least one light quark $q$, 
are modified in medium in the QMC model, but $\Omega, \Omega_c$, and $\Omega_b$ 
are not modified in the QMC model.
We remind that some inputs are updated from those in Refs.~\cite{Saito:2005rv,Krein:2017usp} 
based on the data~\cite{PDG}.
For the recent data for $\Sigma_b$, see Ref.~\cite{Aaij:2018tnn}, which give the 
averaged mass of $m_{\Sigma_b} = 5813.1$ MeV, to be consistent with the value extracted 
from Ref.~\cite{PDG}.
}
\label{bagparambc}
\vspace{1ex}
\begin{tabular}{c|cccccc|ccccc}
\hline
\hline
$B(q_1,q_2,q_3)$ &$z_B$ &$m_B$ &$R_B$ &$x_1$ &$x_2$ &$x_3$ &$m_B^*$ &$R_B^*$ 
&$x^*_1$ &$x^*_2$ &$x^*_3$\\
\hline
$N(qqq)$           &3.295 &939.0  &0.800 &2.052 &2.052 &2.052   & 754.5 &0.786 &1.724 &1.724 &1.724\\
\hline
$\Lambda(uds)$     &3.131 &1115.7 &0.806 &2.053 &2.053 &2.402   & 992.7 &0.803 &1.716 &1.716 &2.401\\
$\Sigma(qqs)$      &2.810 &1193.1 &0.827 &2.053 &2.053 &2.409   &1070.4 &0.824 &1.705 &1.705 &2.408\\
$\Xi(qss)$         &2.860 &1318.1 &0.820 &2.053 &2.406 &2.406   &1256.7 &0.818 &1.708 &2.406 &2.406\\
$\Omega(sss)$      &1.930 &1672.5 &0.869 &2.422 &2.422 &2.422   &---    &---   &---   &---   &---  \\
\hline
$\Lambda_c(udc)$   &1.642 &2286.5 &0.854 &2.053 &2.053 &2.879   &2164.2 &0.851 &1.691 &1.691 &2.878\\
$\Sigma_c(qqc)$    &0.903 &2453.5 &0.892 &2.054 &2.054 &2.889   &2331.8 &0.889 &1.671 &1.671 &2.888\\
$\Xi_c(qsc)$       &1.445 &2469.4 &0.860 &2.053 &2.419 &2.880   &2408.3 &0.859 &1.687 &2.418 &2.880\\
$\Omega_c(ssc)$    &1.057 &2695.2 &0.876 &2.424 &2.424 &2.884   &---    &---   &---   &---   &---  \\
\hline
$\Lambda_b(udb)$  &-0.622 &5619.6 &0.930 &2.054 &2.054 &3.063   &5498.5 &0.927 &1.651 &1.651 &3.063\\
$\Sigma_b(qqb)$   &-1.554 &5813.4 &0.968 &2.054 &2.054 &3.066   &5692.8 &0.966 &1.630 &1.630 &3.066\\
$\Xi_b(qsb)$      &-0.785 &5793.2 &0.933 &2.054 &2.441 &3.063   &5732.7 &0.931 &1.649 &2.440 &3.063\\
$\Omega_b(ssb)$   &-1.327 &6046.1 &0.951 &2.446 &2.446 &3.065   &---    &---   &---   &---   &---  \\
\hline
\hline   
\end{tabular}
\end{center}
\end{table}
%

One can notice a few things easily in table~\ref{bagparambc}: (i) the parameter 
$z_B$ decreases as the vacuum mass of the baryon increases, 
(ii) the in-medium bag radius $R^*_B$ of the baryon $B$ at $\rho_0$ decreases than 
the corresponding vacuum value, and the decreasing ratio becomes smaller as the vacuum baryon mass 
value increases, and (iii) the lowest mode bag eigenvalues decrease at $\rho_0$, and the 
decreasing magnitude is larger for the light quarks, but tiny for the heavier quarks.
Note that, the bag radius is not the physical observable, and one must 
calculate the baryon radius using the corresponding quark wave function. 
In fact, such calculation shows that the slight increase of the in-medium radius. 
(See table 2 in Ref.~\cite{Saito:1996sf}.)

\begin{figure}
\centering
\includegraphics[scale=0.3]{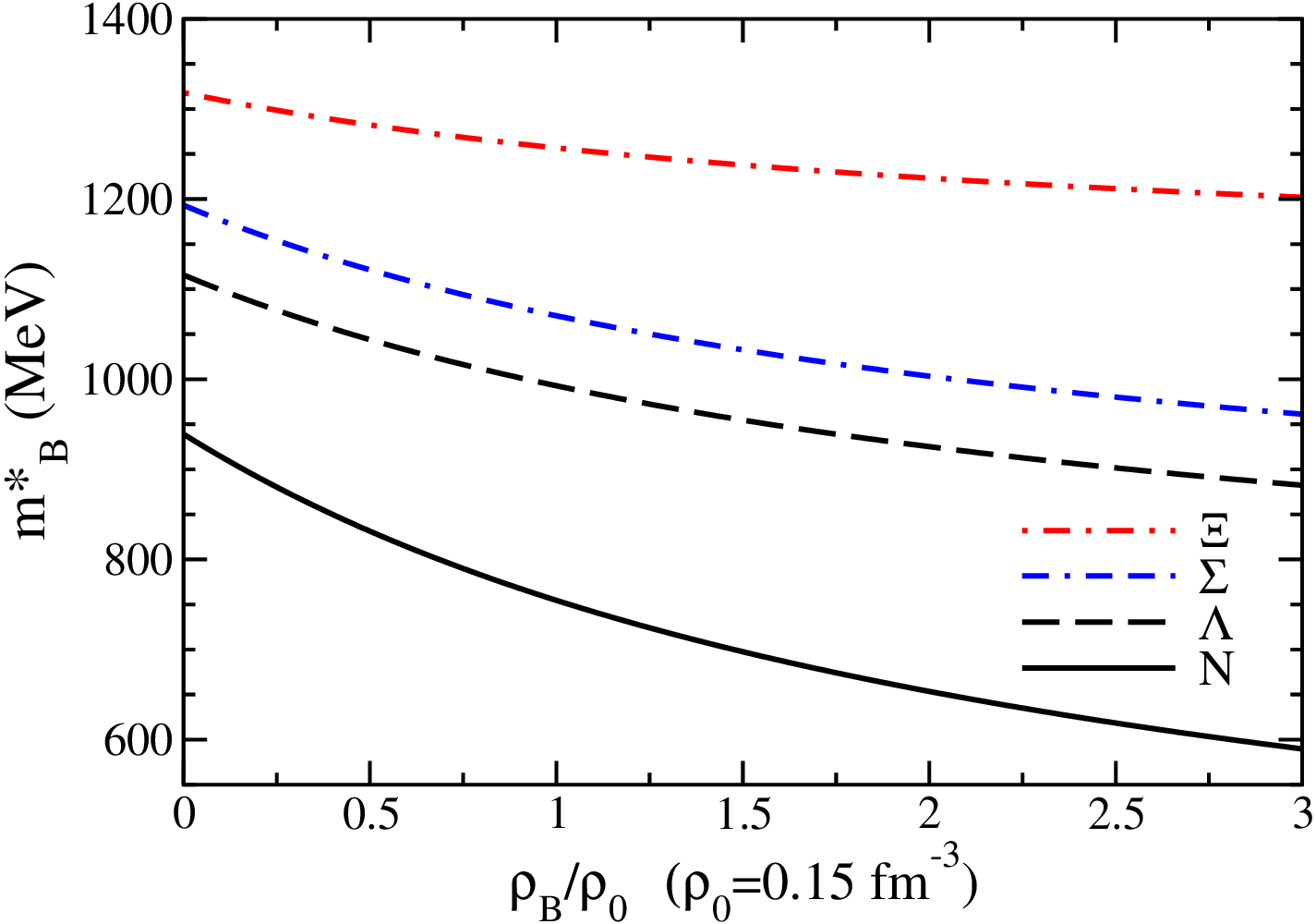}
\vspace{7ex}
\\
\includegraphics[scale=0.3]{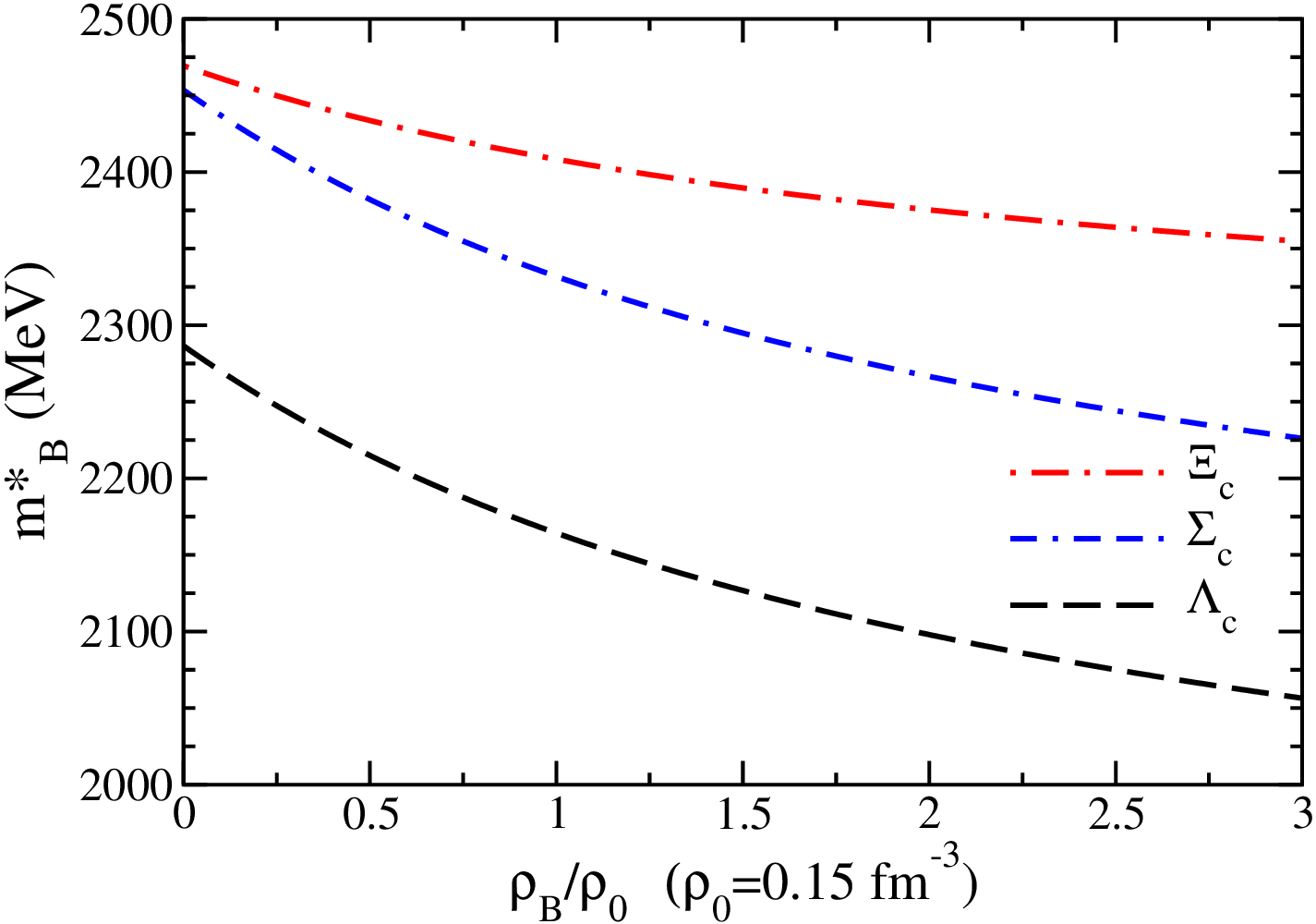}
\hspace{2ex}
\includegraphics[scale=0.3]{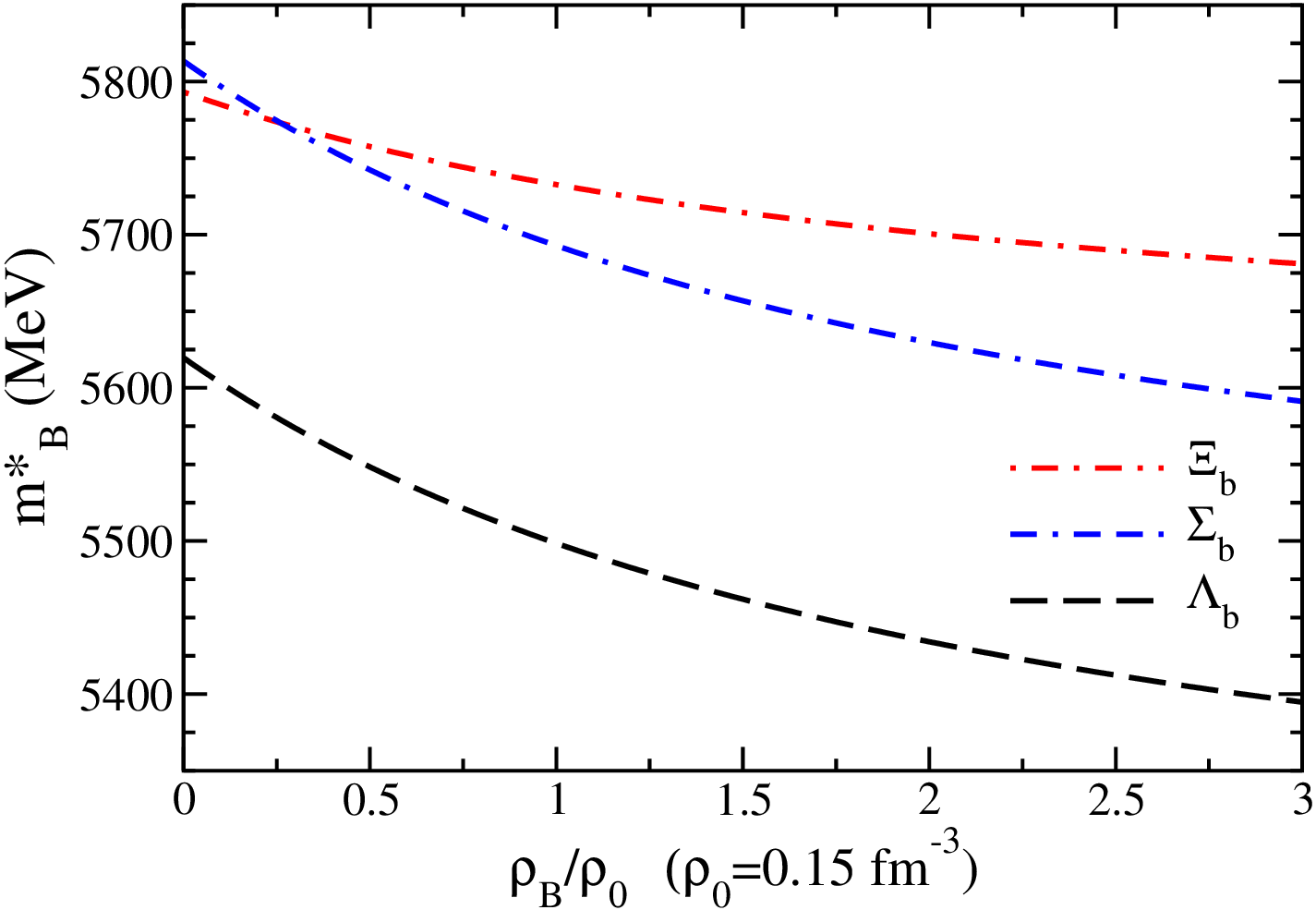}
\caption{\label{fig:baryonmass} 
Density dependence of Lorentz-scalar baryon effective masses in symmetric nuclear matter.
\vspace{5ex}
}
\vspace{2ex}
\end{figure}

\begin{figure}
\centering
\includegraphics[scale=0.48]{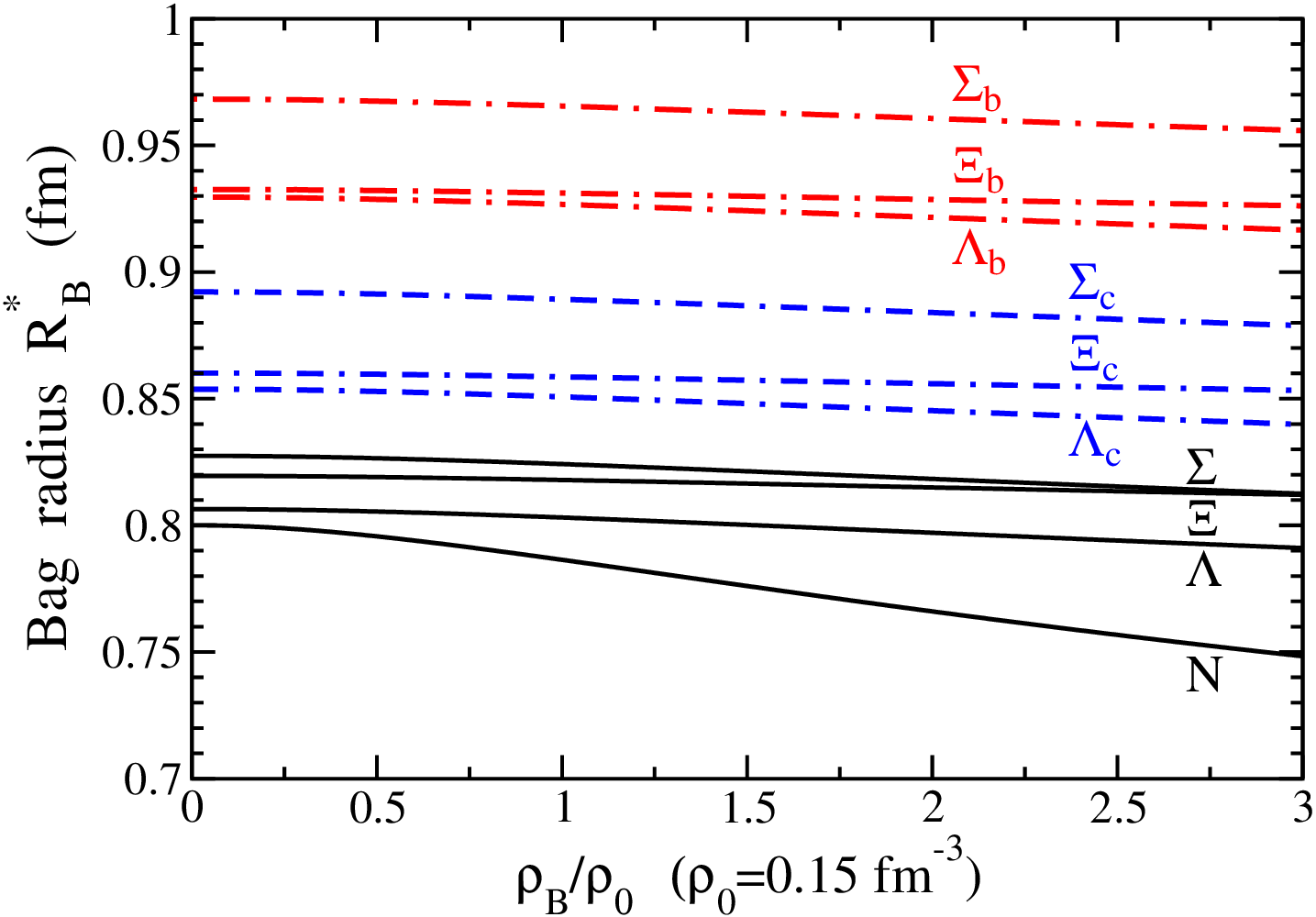}
\caption{\label{fig:BagRB} 
Density dependence of in-medium bag radii in symmetric nuclear matter.
}
\vspace{2ex}
\end{figure}

\begin{figure}
\centering
\includegraphics[scale=0.3]{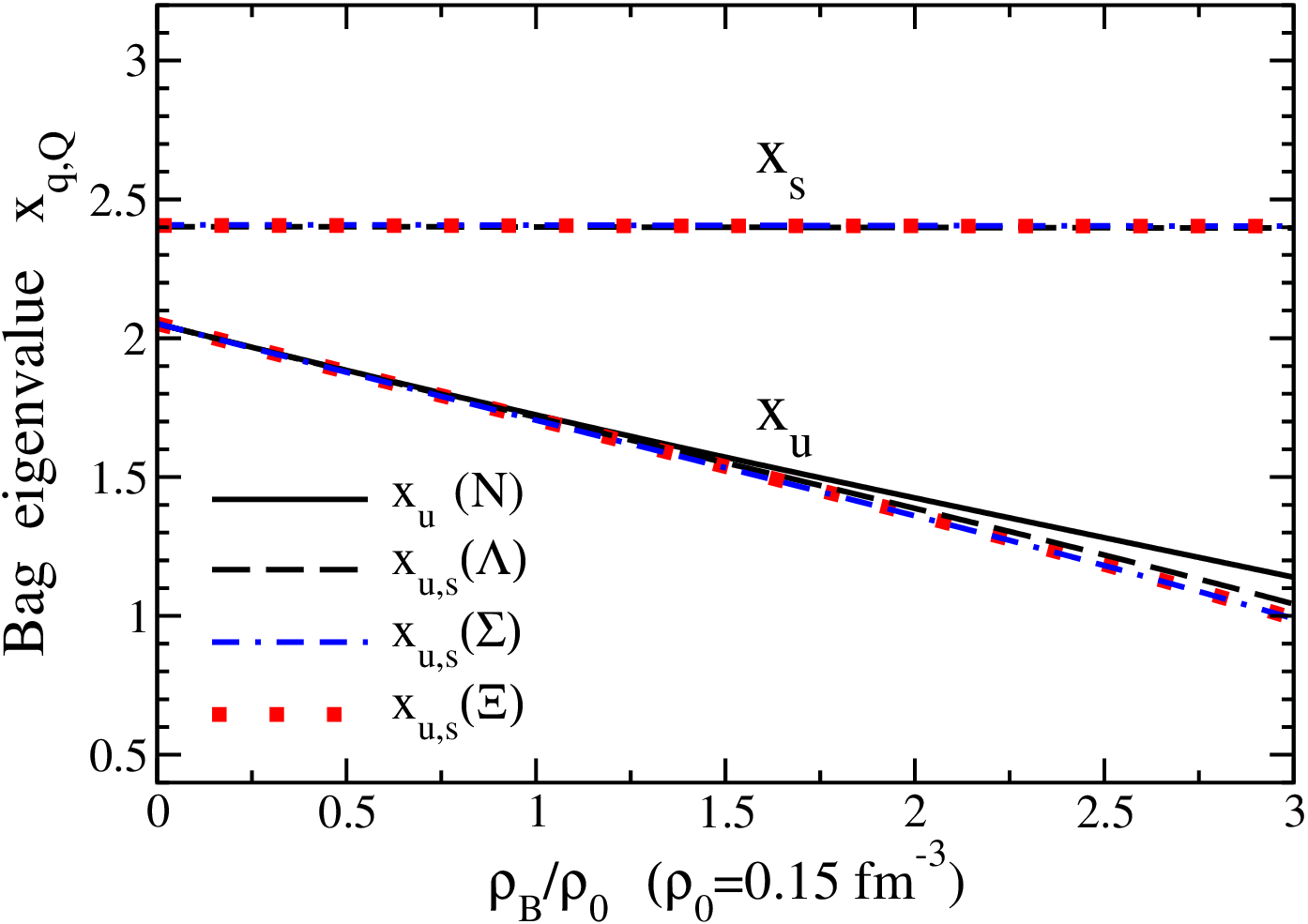}
\vspace{7ex}
\\
\includegraphics[scale=0.3]{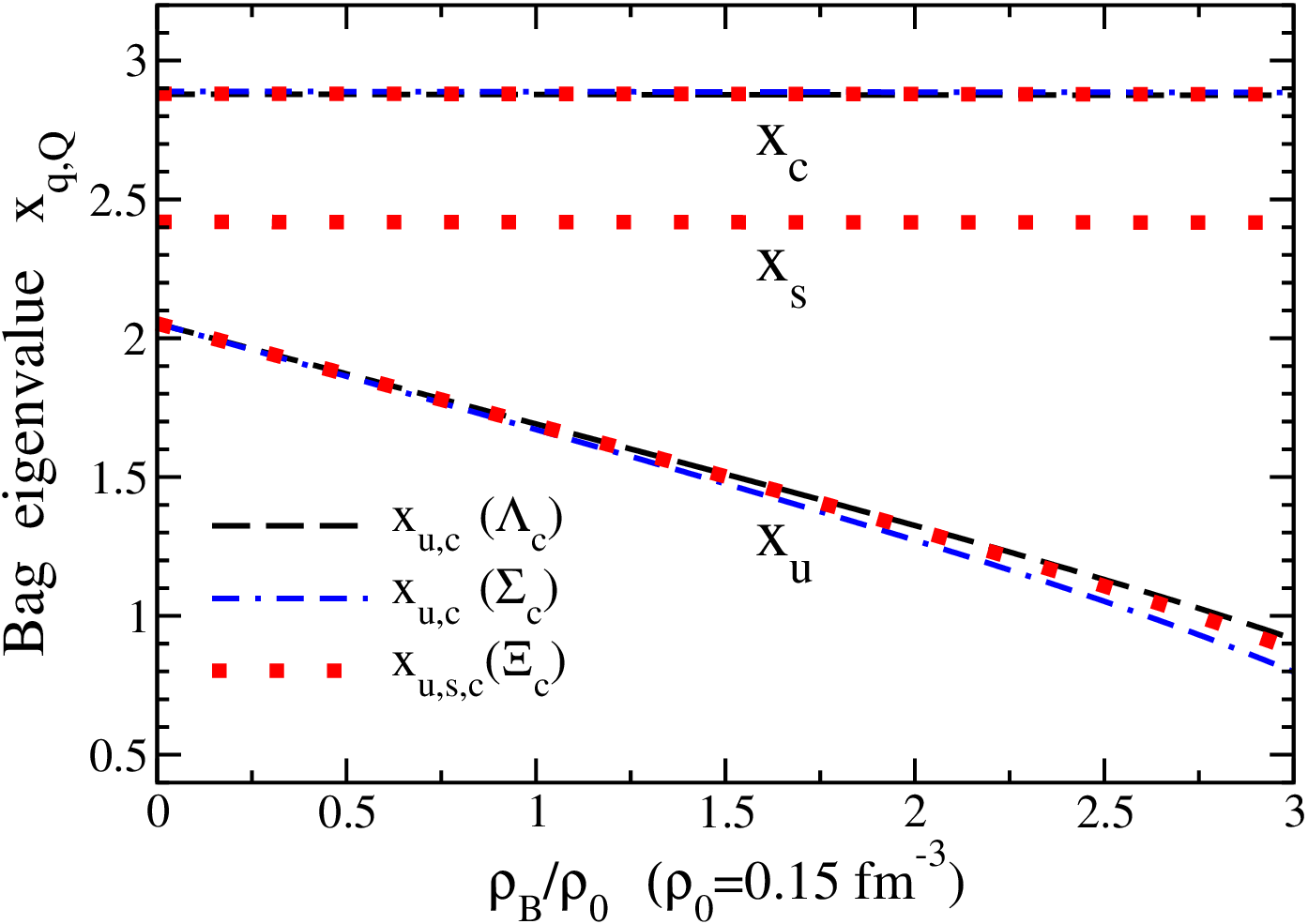}
\hspace{2ex}
\includegraphics[scale=0.3]{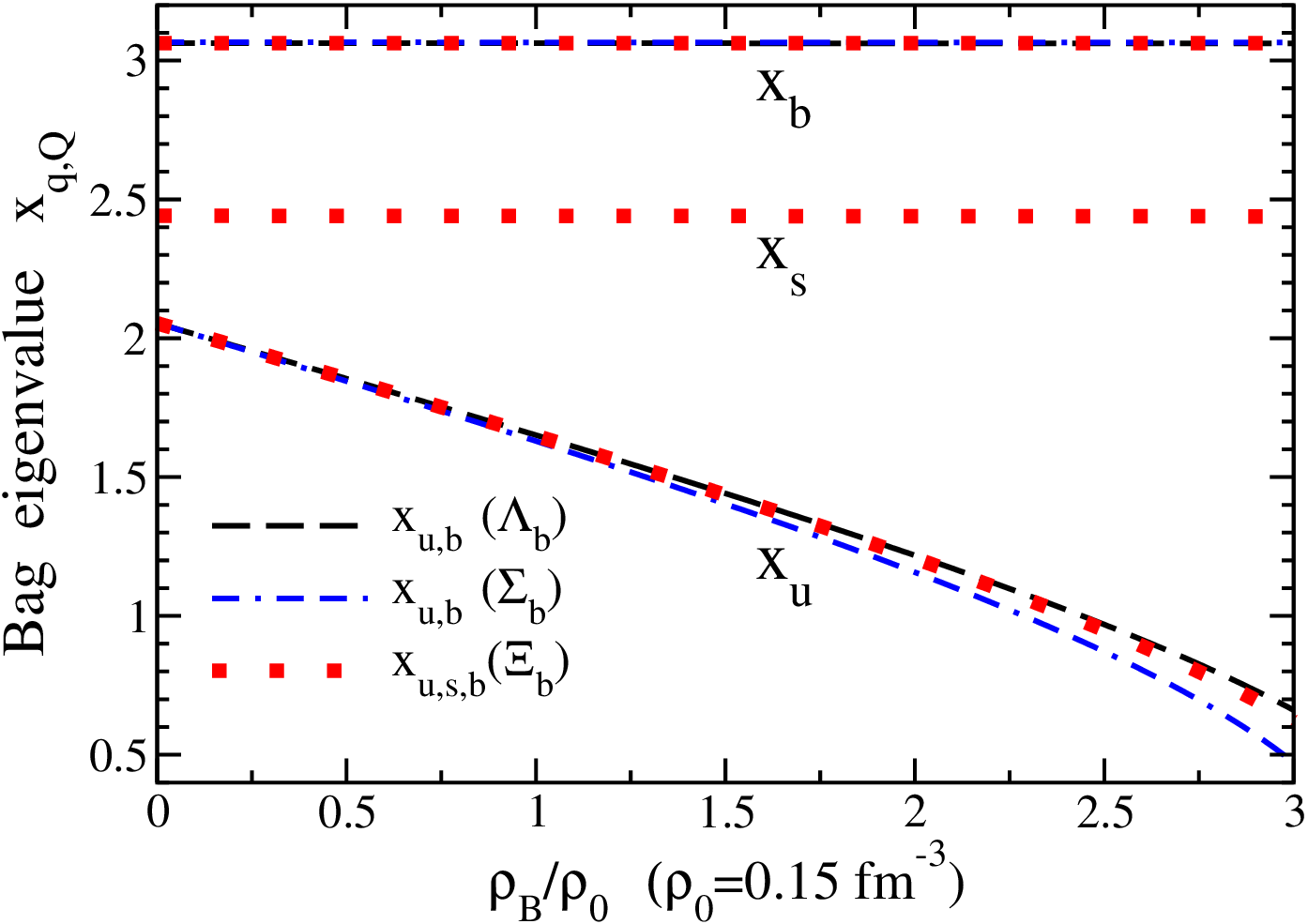}
\caption{\label{fig:eigens} 
Density dependence of the lowest mode bag eigenvalues in symmetric nuclear matter.
}
\vspace{2ex}
\end{figure}

In Figs.~\ref{fig:baryonmass}, \ref{fig:BagRB}, and~\ref{fig:eigens} 
we show respectively the density dependence of the Lorentz-scalar effective baryon masses, 
in-medium bag radii, and the lowest mode bag eigenvalues.
In Figs.~\ref{fig:baryonmass} and~\ref{fig:eigens} each panel respectively shows for 
nucleon and strange baryons (top panel), for charm baryons (bottom-left panel), 
and for bottom baryons (bottom-right panel).

For the Lorentz-scalar effective masses shown in Fig.~\ref{fig:baryonmass}, one can notice 
a very interesting feature for the bottom baryons (bottom-right panel).
The Lorentz-scalar effective mass of $\Sigma_b$ becomes smaller 
than that of $\Xi_b$, namely  $m^*_{\Sigma_b} < m^*_{\Xi_b}$ 
at baryon density range larger than about $0.3 \rho_0$, 
although vacuum masses satisfy $m_{\Sigma_b} > m_{\Xi_b}$~\cite{PDG} 
(see table~\ref{bagparambc}). This is indeed interesting, and can 
be understood as follows. The $\Sigma_b$ baryon contains two light quarks, 
while the $\Xi_b$ baryon contains one. Because the light quark 
condensates are much more sensitive to the nuclear density change 
than those of the strange, charm and bottom 
quark ones, one can expect that the partial restoration of chiral symmetry 
to take place faster for $\Sigma_b$ than $\Xi_b$ as increasing the 
nuclear density. Or in the QMC model 
picture, since the scalar potential is roughly proportional to the 
number of the valence light 
quarks~\cite{Saito:2005rv,Tsushima:1997cu,Tsushima:2002cc}, 
the Lorentz-scalar effective mass of $\Sigma_b$  
decreases faster than that of $\Xi_b$ as increasing the 
nuclear matter density.

The result of the reverse in the Lorentz-scalar effective masses of $\Sigma_b$ and $\Xi_b$  
is one of the main predictions of this article.
We must seek how this interesting prediction possibly be connected 
with experimental observables. 
This would give very important information on the dynamical 
symmetry breaking and the partial restoration of chiral (dynamical) symmetry.
However, the story is not that straightforward,  
since the baryons (light quarks) also feel repulsive Lorentz-vector potentials 
in addition to the attractive Lorentz-scalar potentials.
Thus, we must take into account the effects of the repulsive vector potentials 
for considering more realistic/practical experimental situations, 
and we will study this later.

Concerning the in-medium bag radii shown in Fig.~\ref{fig:BagRB}, 
one can notice that all the in-medium bag radii decrease as increasing 
the nuclear matter density.
In particular, the decrease for the nucleon case is the largest.

As for the lowest mode bag eigenvalues shown in Fig.~\ref{fig:eigens}, they also decrease 
as increasing the nuclear matter density, particularly noticeable for the light quarks, 
but tiny decreases for the heavier quarks.

In connection with the Lorentz-scalar effective baryon masses shown in Fig.~\ref{fig:baryonmass},
it has been found that the function $C_B({\sigma})\, 
(B = N,\Lambda,\Sigma,\Xi,\Lambda_c,\Sigma_c,\Xi_c,\Lambda_b,\Sigma_b,\Xi_b)$
(see Eq.~(\ref{Ssigma}) and above), 
can be parameterized as a linear
form in the $\sigma$ field, $g_{\sigma}{\sigma}$, for a practical
use~\cite{Guichon:1995ue,Saito:1996sf,Tsushima:1997cu}:
\begin{equation}
C_B ({\sigma}) = 1 - a_B
\times (g_{\sigma} {\sigma}),\hspace{1em}
(B = N,\Lambda,\Sigma,\Xi,\Lambda_c,\Sigma_c,\Xi_c,\Lambda_b,\Sigma_b,\Xi_b).
\label{cynsigma}
\end{equation}
The values obtained for $a_B$ are listed in table~\ref{slope}.
This parameterization works very well up to
about three times of normal nuclear matter density $3 \rho_0$.
Then, the effective mass of baryons $B$ in nuclear matter
is well approximated by:
\bge
m^*_B \simeq m_B - \dfrac{n_q}{3} g_\sigma 
\left[1-\dfrac{a_B}{2}(g_\sigma {\sigma})\right]\sigma,
\hspace{2ex}(B = N,\Lambda,\Sigma,\Xi,\Lambda_c,\Sigma_c,\Xi_c,\Lambda_b,\Sigma_b,\Xi_b),
\label{Mstar}
\ene
with $n_q$ being the valence light-quark number in the baryon $B$.
See Eqs.~(\ref{effnmass}) and~(\ref{effymass}) to compare with 
$g^{N,Y}(\sigma)$ and the above expression.
For the $\Sigma_b$ and $\Xi_b$ baryons, $n_q$ are respectively two and one  
in Eq.~(\ref{Mstar}) with $a_{\Sigma_b} \simeq a_{\Xi_b}$ 
from table~\ref{slope}. Then, one can confirm that the decrease in 
the Lorentz-scalar effective mass for $\Sigma_b$ is larger than that for $\Xi_b$ 
as increasing the nuclear matter density, or as increasing the $\sigma$ mean field.

%
\begin{table}
\begin{center}
\caption{Slope parameters, $a_B\,\,
(B = N,\Lambda,\Sigma,\Xi,\Lambda_c,\Sigma_c,\Xi_c,\Lambda_b,\Sigma_b,\Xi_b)$.
Note that the tiny differences in values of $a_B$ from those in  
Refs.~\cite{Saito:2005rv,Krein:2017usp}, are due to the differences 
in the number of data points for calculating $a_B$, 
but such differences in $a_B$ give negligible effects.   
\vspace{1ex}
}
\label{slope}
\begin{tabular}[t]{c|c||c|c||c|c}
\hline
$a_B$ &$\times 10^{-4}$ MeV$^{-1}$ &$a_B$ &$\times 10^{-4}$ MeV$^{-1}$ 
&$a_B$ &$\times 10^{-4}$ MeV$^{-1}$\\
\hline
$a_N$         &9.1  &--- &---                &--- &--- \\
$a_\Lambda$   &9.3  &$a_{\Lambda_c}$ &9.9    &$a_{\Lambda_b}$ &10.8 \\
$a_{\Sigma}$  &9.6  &$a_{\Sigma_c}$  &10.3   &$a_{\Sigma_b}$  &11.2 \\
$a_{\Xi}$     &9.5  &$a_{\Xi_c}$     &10.0   &$a_{\Xi_b}$     &10.8 \\
\hline
\end{tabular}
\end{center}
\end{table}
%

\begin{figure}[htb]
\centering
\vspace{1.5ex}
\includegraphics[scale=0.3]{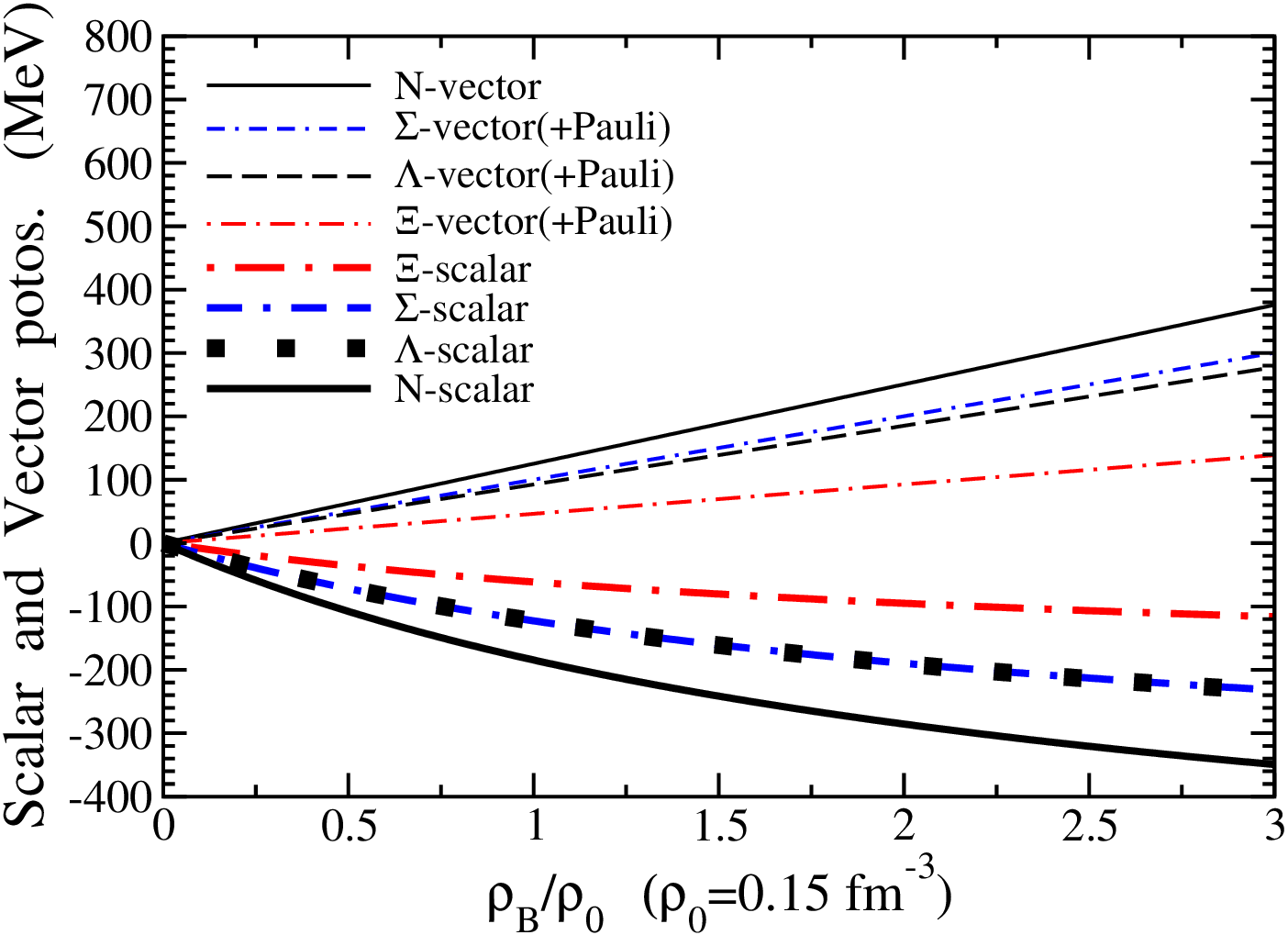}
\vspace{7ex}
\\
\includegraphics[scale=0.3]{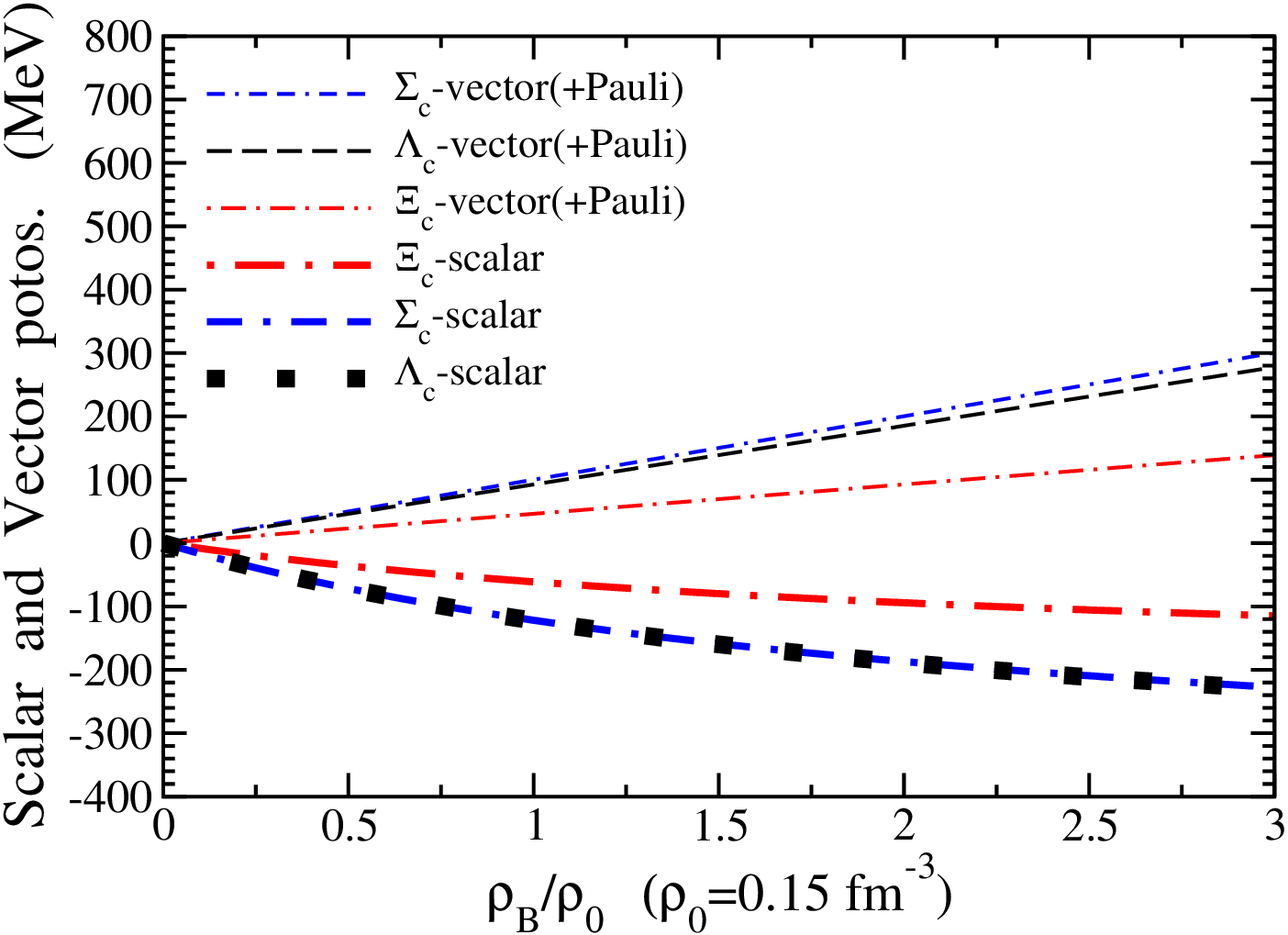}
\hspace{2ex}
\includegraphics[scale=0.3]{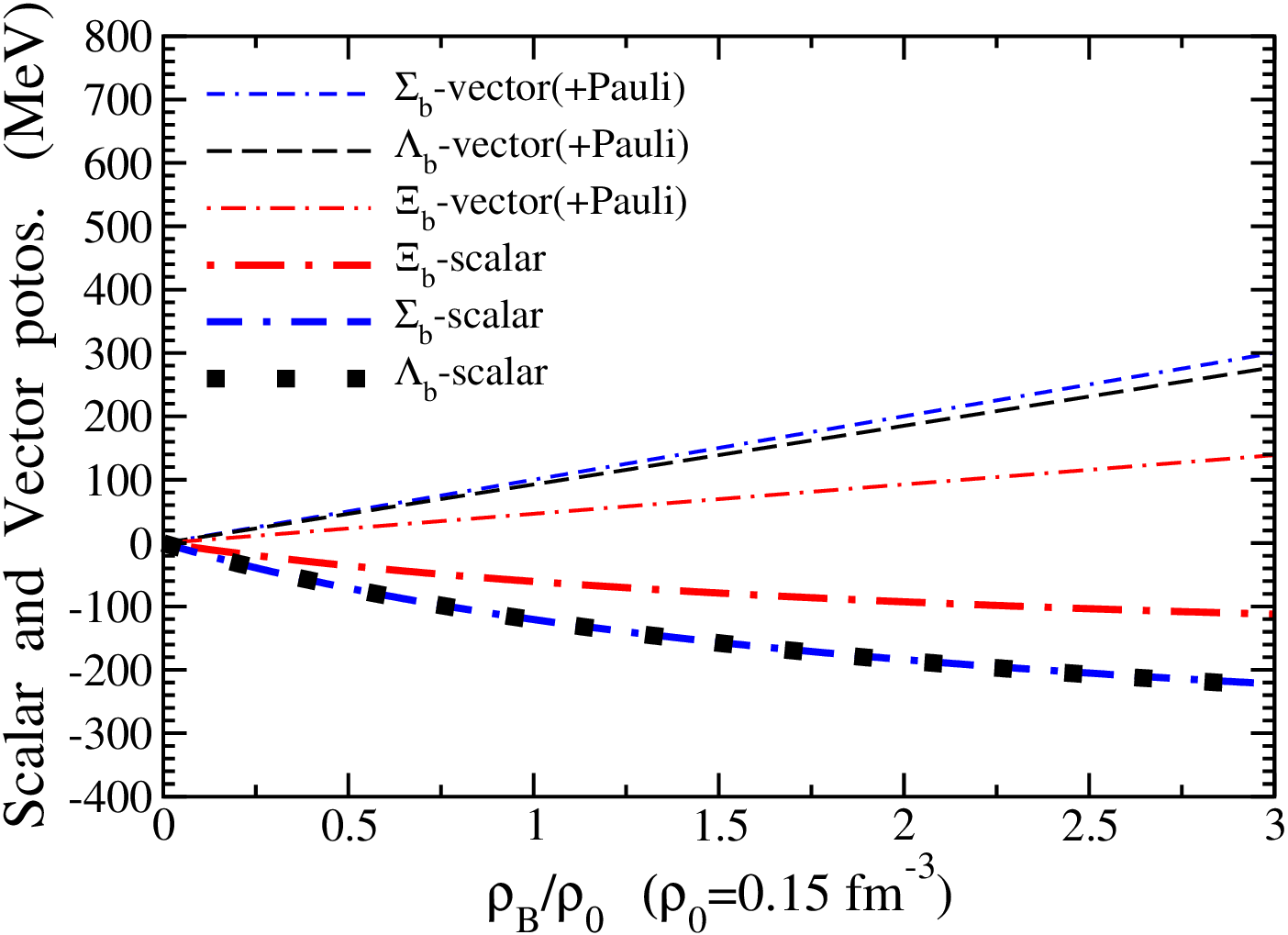}
\caption{\label{fig:SVpot} 
Attractive Lorentz-scalar and repulsive Lorentz-vector potentials   
of baryons in symmetric nuclear matter. 
In figures ``vector+Pauli'' denote that the effective Pauli potentials are added, 
and ``potos.'' in each vertical axis is the abbreviation for ``potentials''. 
}
\vspace{2ex}
\end{figure}

\begin{figure}
\centering
\vspace{2.5ex}
\includegraphics[scale=0.3]{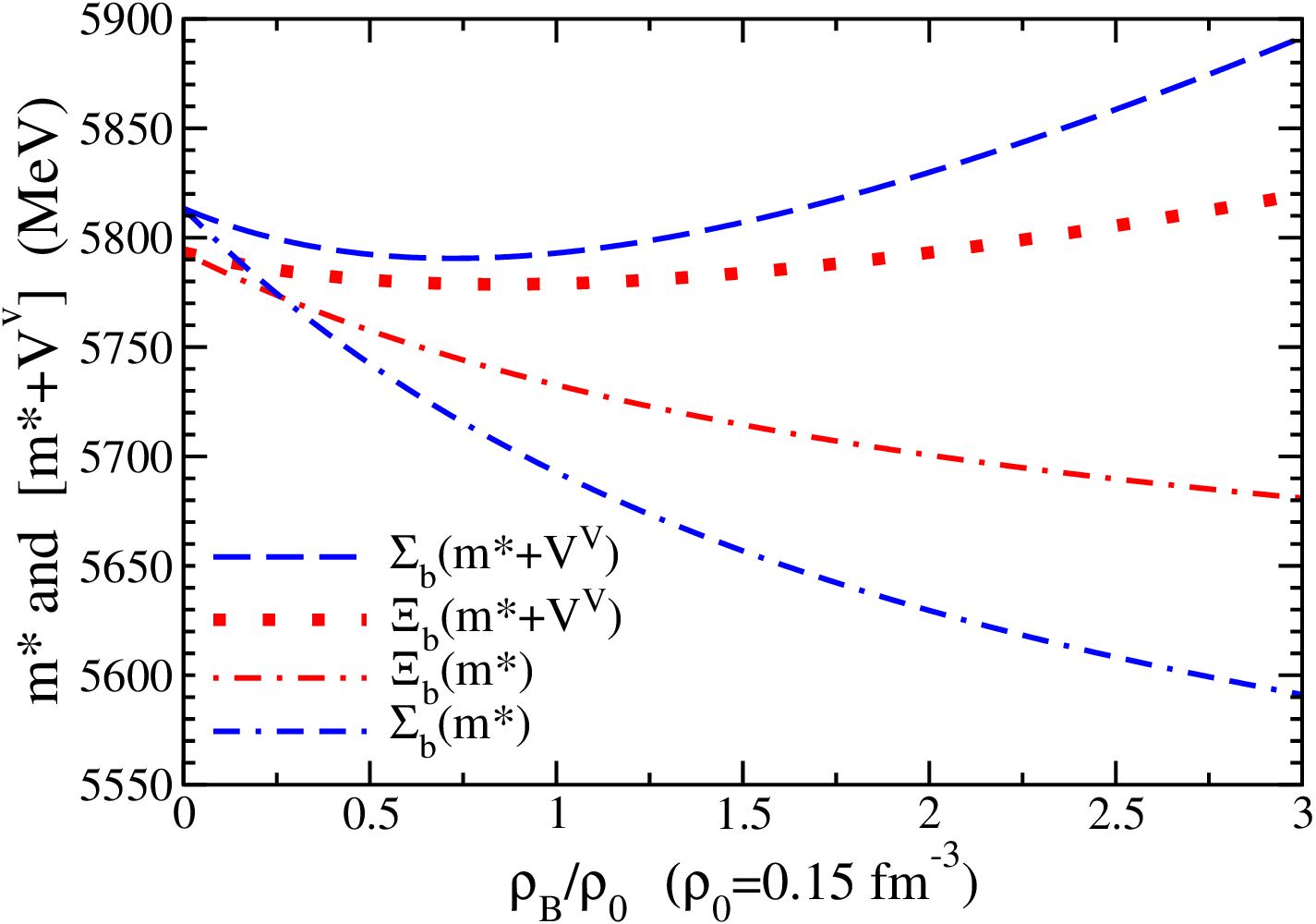}
\hspace{1ex}
\includegraphics[scale=0.3]{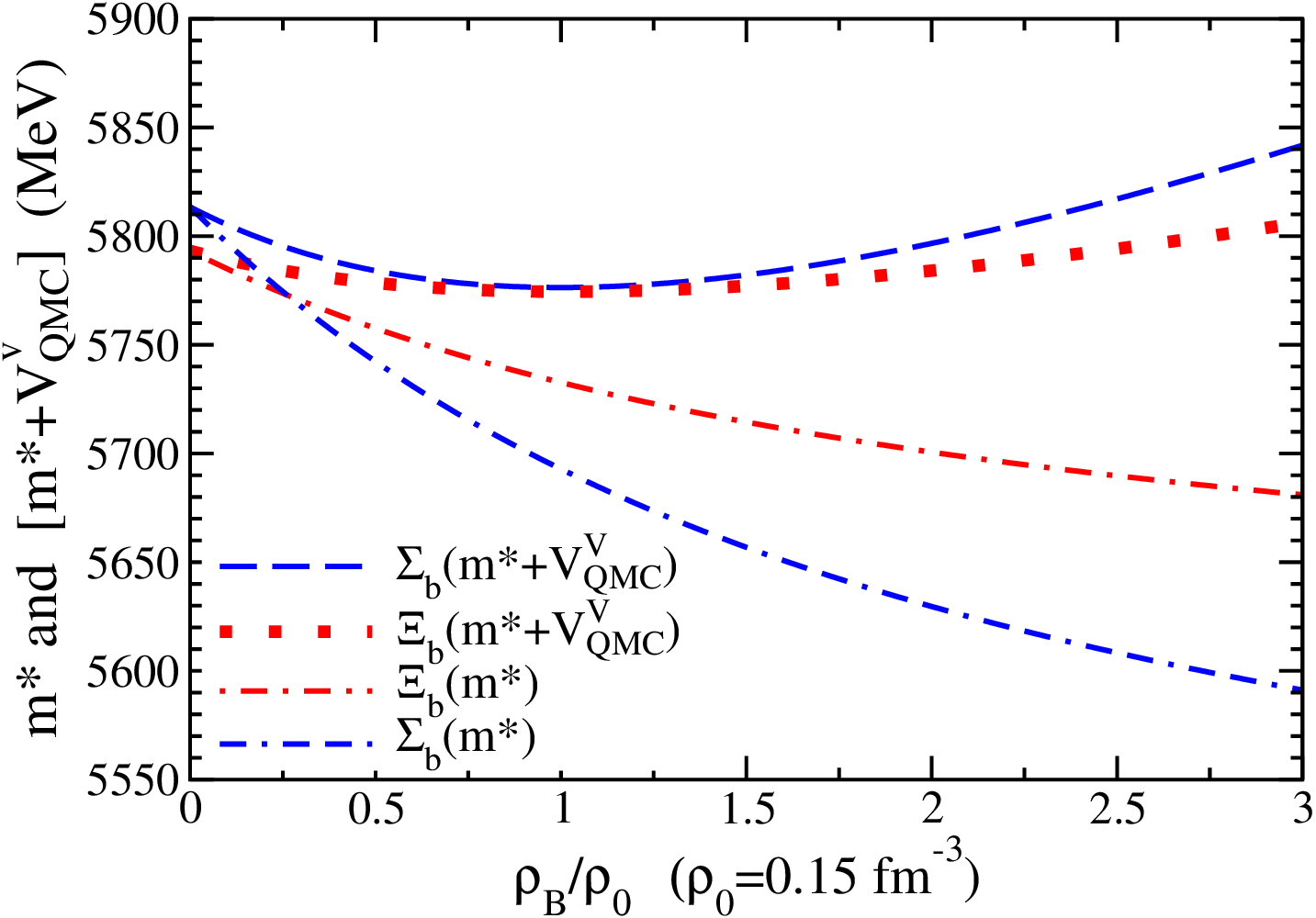}
\caption{\label{fig:SbXbEnergy} 
Effective masses and excitation energies (total potentials) of $\Sigma_b$ and $\Xi_b$ baryons for the two cases of the vector potentials, with including the Pauli potentials (left panel) and without 
(right panel).
}
\vspace{2ex}
\end{figure}

To analyze more carefully the interesting findings for the $\Sigma_b$ and $\Xi_b$ baryon
Lorentz-scalar effective masses, we next discuss the ``excitation energies'' 
of baryons, to study the total energies (potentials) in a nonrelativistic sense, 
the Lorentz-scalar plus Lorentz-vector potentials focusing 
on the $\Sigma_b$ and $\Xi_b$ baryons.
First, results for the attractive scalar and repulsive vector potentials separately, are shown in Fig.~\ref{fig:SVpot}, for nucleon and strange baryons (top panel), 
charm baryons (bottom-left panel), 
and bottom baryons (bottom-right panel).
For the repulsive vector potentials, we show here only one case, 
the one including the effective ``Pauli potentials'' introduced 
in Eq.~(\ref{Pauli}), denoted by ``vector(+Pauli)''. 
One can see the similarity in the amounts of the scalar and vector(+Pauli) potentials 
among the corresponding strange, charm and bottom sector baryons, 
namely among those three baryons in each brackets,  
($\Lambda,\Lambda_c,\Lambda_b$), ($\Sigma,\Sigma_c,\Sigma_b$) 
and ($\Xi,\Xi_c,\Xi_b$).

Now we show in Fig.~\ref{fig:SbXbEnergy} Lorentz-scalar effective masses and  
excitation energies (total energies), Lorentz-scalar effective masses plus vector potentials 
for the two cases of the vector potentials focusing on $\Sigma_b$ and $\Xi_b$.
The left panel is the case with the Pauli potentials, while the right panel is without 
the Pauli potentials. Recall that, because the mass difference between 
the $\Lambda_b - \Sigma_b$ system is much larger than that for 
the $\Lambda - \Sigma$ and $\Lambda_c - \Sigma_c$ systems, 
it is expected that the effective Pauli potentials should be smaller for 
the $\Lambda_b, \Sigma_b$ and $\Xi_b$ baryons than the corresponding 
strange and charm sector baryons. 
Thus, one can regard the more realistic case when we consider 
without the Pauli potentials, shown in the right panel of Fig.~\ref{fig:SbXbEnergy}.

We discuss separately the two cases of the vector potentials.
First, for the case with the Pauli potentials shown in the left panel 
of Fig.~\ref{fig:SbXbEnergy}, the excitation energies (total potentials) for the 
$\Sigma_b$ and $\Xi_b$ never reverse in magnitudes, and always the excitation 
energy of $\Sigma_b$ is larger than that for $\Xi_b$. 
The smallest excitation energy difference is about a few tens of MeV, and it is larger 
for $\Sigma_b$. For the nuclear matter density larger than around $\rho_0$, 
the difference in the excitation energies increases.

Next, for the case without the Pauli potentials, which 
may be expected to be more realistic, is shown in the right panel of Fig.~\ref{fig:SbXbEnergy}. 
Interestingly, in the 
nuclear matter density range $0.5 \rho_0 < \rho_B < 1.5 \rho_0$, the two 
excitation energies for $\Sigma_b$ and $\Xi_b$ are nearly degenerate.
This means that $\Sigma_b$ and $\Xi_b$ can be produced at rest with the nearly same costs  
of energies. This may imply the emergence of many interesting phenomena,  
for example, in heavy ion reactions and reactions in the systems of dense nuclear medium, 
such as in a deep core of neutron (compact) star.

The results shown in Fig.~\ref{fig:SbXbEnergy} suggest that  
the two different types of the vector potentials may possibly be distinguished, 
and give important information on the dynamical symmetry breaking 
and partial restoration of chiral symmetry,  
by studying the heavy bottom baryon properties in medium.
For proving these suggestions, we have to seek what kind of experiments can be made 
to get a clue, in particular, for the Lorentz-scalar effective masses of $\Sigma_b$ and $\Xi_b$.
It might be very interesting to measure the valence quark (parton) distributions 
of $\Sigma_b$ and $\Xi_b$ in medium, since the supports of the parton distributions  
of these baryons reflect their excitation energies. Other possibility 
may be to measure the strangeness-changing semi-leptonic weak decay 
of $\Xi_b \to \Sigma_b$ in medium, which again reflects the excitation energy difference  
of them in medium.

\section{Summary and discussion}
\label{summary}

In this article we have completed the study of 
baryon properties in symmetric nuclear matter 
in the quark-meson coupling model, for the low-lying strange, 
charm, and bottom baryons which contain at least one light quark.
We have presented the density dependence of the Lorentz-scalar effective masses, 
bag radii, the lowest mode bag eigenvalues, and vector potentials for the baryons.

We predict that the Lorentz-scalar effective mass of $\Sigma_b$ becomes 
smaller than that of $\Xi_b$ in the nuclear matter density    
range larger than $\simeq 0.3 \rho_0$ ($\rho_0 = 0.15$ fm$^{-3}$), 
while in vacuum the mass of $\Sigma_b$ is larger than that of $\Xi_b$.
We also give parametrization for the Lorentz-scalar effective masses of the baryons treated 
in this article as a function of the scalar mean field for a convenient use.  

We have further studied the effects of the two different repulsive Lorentz-vector 
potentials to estimate the excitation (total) 
energies focusing on $\Sigma_b$ and $\Xi_b$ baryons.
In the case without the effective Pauli potentials, 
which is expected to be more realistic, the excitation energies 
for the $\Sigma_b$ and $\Xi_b$ baryons are predicted to be nearly degenerate 
in the nuclear matter density range about [$0.3 \rho_0, 1.5 \rho_0$]. 
Thus, the production of $\Sigma_b$ and $\Xi_b$ baryon cost nearly the 
same energies at rest in this nuclear matter density range, and this may imply 
many interesting phenomena in heavy ion collisions, and reactions involving them 
in a deep core of neutron (compact) star.

To make possible connections of the findings for the Lorentz-scalar effective masses and/or 
excitation energies of $\Sigma_b$ and $\Xi_b$ baryons with experimental observables, 
we need to seek relevant experimental methods and situations.
It might be very interesting to measure the valence quark (parton) distributions 
of $\Sigma_b$ and $\Xi_b$ in medium, since the supports of the parton distributions  
of these baryons reflect their excitation energies. Other possibility 
may be to measure the strangeness-changing semi-leptonic weak decay   
of $\Xi_b \to \Sigma_b$ in medium, which again reflects the excitation 
energy difference of them in medium.

In conclusion, studies of heavy baryon properties, in particular $\Sigma_b$ and $\Xi_b$ 
baryons in nuclear medium, can provide us with very interesting and important information 
on the dynamical symmetry breaking and partial restoration of chiral symmetry, 
as well as the roles of the light quarks in medium.

\vspace{1ex}
\noindent
{\bf Acknowledgments}\\
This work was partially supported by
the Conselho Nacional de Desenvolvimento Cient\'{i}fico e Tecnol\'{o}gico - CNPq  
Grants, No.~400826/2014-3 and No.~308088/2015-8, and was also part of the projects, 
Instituto Nacional de Ci\^{e}ncia e Tecnologia - Nuclear Physics and Applications (INCT-FNA), 
Brazil, Process No. 464898/2014-5, and FAPESP Tem\'{a}tico, 
Brazil, Process No. 2017/05660-0.


\end{document}